\documentclass[aps,onecolumn,groupedaddress,nofootinbib] {revtex4-1}
\usepackage{amsmath}
\usepackage{amsfonts}
\usepackage{amssymb}
\usepackage{graphicx}
\usepackage{caption}
\usepackage{subcaption}
\usepackage{float}
\usepackage{braket}
\allowdisplaybreaks
\newcommand {\bp}{\begin{pmatrix}}
\newcommand {\ep}{\end{pmatrix}}
\newcommand{\be}{\begin{equation}} \newcommand{\ee}{\end{equation}}
\newcommand{\bea}{\begin{eqnarray}}\newcommand{\eea}{\end{eqnarray}}

\DeclareMathOperator{\sech}{sech}
\DeclareMathOperator{\sgn}{sgn}
\begin{document}

\title{Non-linear Schr$\ddot{o}$dinger equation with time-dependent balanced 
loss-gain and space-time modulated non-linear interaction}
\author{Supriyo Ghosh}
\email[]{supriyoghosh711@gmail.com}
\author{Pijush K. Ghosh}
\email[Corresponding Author: ]{pijushkanti.ghosh@visva-bharati.ac.in}
\affiliation{ Department of Physics, Siksha-Bhavana, Visva-Bharati, 
Santiniketan, PIN 731 235, India.}
\date{\today}

\begin{abstract}
We consider a class of one dimensional vector Non-linear Schr$\ddot{o}$dinger
Equation(NLSE) in an external complex potential with Balanced Loss-Gain(BLG)
and Linear Coupling(LC) among the components of the Schr$\ddot{o}$dinger field.
The solvability of the generic system is investigated for various combinations
of time modulated LC and BLG terms, space-time dependent strength of the nonlinear 
interaction and complex potential. We use a non-unitary transformation followed 
by a reformulation of the differential equation in a new coordinate system to map 
the NLSE to solvable equations. Several physically
motivated examples of exactly solvable systems are presented for various
combinations of LC and BLG, external complex potential and nonlinear
interaction. Exact localized nonlinear modes with spatially constant phase
may be obtained for any real potential for which the corresponding linear
Schr$\ddot{o}$dinger equation is solvable. A method based on supersymmetric
quantum mechanics is devised to construct exact localized nonlinear modes
for a class of complex potentials. The real superpotential corresponding
to any exactly solved linear Schr$\ddot{o}$dinger equation  may be used
to find a complex-potential for which exact localized nonlinear modes
for the NLSE can be obtained. The solutions with singular phases are
obtained for a few complex potentials.
\end{abstract}
\maketitle

\section{Introduction}
The Non-linear Schr$\ddot{o}$dinger equation(NLSE) appears in the context of mathematical 
modelling of many physical phenomena in different domains of science, like optics in 
nonlinear media \cite{Kivshar,Serkin,Serkin1}, Bose-Einstein Condensates(BEC)\cite{Dalfovo,
Pitaevskii1,Kevrekidis,Kengne}, plasma physics\cite{Dodd},  gravity waves\cite{Trulsen},
Bio-molecular dynamics\cite{Davydov} etc. NLSE exhibits
soliton\cite{Zabusky,Newell,Zakharov3,Infeld} solutions and is an integrable system with a very 
rich mathematical structure\cite{Sulem,Bourgain,Ablowitz,Lamb}. The quantized NLSE describes a 
hardcore Bose-gas with delta-function interaction for which various physical properties 
including correlations functions can be computed analytically\cite{korepin}.
Several generalisations of the NLSE have been considered to model different emerging phenomena
in physics over the last few decades. For example, the NLSE with an external confining potential
is known as Gross-Pitaevskii equation\cite{Pitaevskii1} and is relevant in the context of BEC. 
The investigations on collapse and growth of quasi-condensates in  NLSE with harmonic confinement 
offers a rich variety of mathematical structure\cite{pkg2,pkg3,Wadati}. Localized, unbounded and 
periodic potentials have also been investigated with interesting results. The NLSE with space
\cite{Abdullaev,Sakaguchi,Carpentier,Beitia,Verde,Theocharis,Porter} and time
\cite{Saito,Kevrekidis1,Brazhnyi,Berge,Towers,Kevrekidis2,Abdullaev1,Garcia,Abdullaev2,Montesinos,
Itin,Konotop} modulated nonlinear strengths is also of immense interest in the same context.

The generalizations of NLSE based on underlying ${\cal{PT}}$-symmetry\cite{Benderl} may be broadly 
classified into three different classes: (i) NLSE with ${\cal{PT}}$-symmetric complex confining 
potential, (ii) the non-local NLSE, and (iii) NLSE with LC and BLG terms. Solvable  models of NLSE 
with ${\cal{PT}}$-symmetric complex localized potential\cite{Konotop1,Shi,Ortiz1,Ortiz2,Cruz,Midya,
Midya1,Musslimani} have been considered  which admits soliton solutions. The investigations on 
${\cal{PT}}$-symmetry breaking as a function of the system parameters reveal a host of interesting 
issues including the behaviour of solitons at the exceptional points\cite{Konotop1}. The non-local 
NLSE\cite{Ablowitz1} is integrable and admits bright as well as dark soliton for the same range of 
the nonlinear strength. The non-local vector NLSE is also integrable and amenable to exact 
solutions and real-valued physical quantities\cite{ds-pkg}. Further, it has lead to a new branch of 
integrable models where non-local generalizations of previously known integrable models have been
considered\cite{pkg1,Ablowitz-2,Avadh}.

The central focus of this article is on NLSE with LC and BLG terms for which, in addition to 
real-valued linear and nonlinear couplings, the components of NLSE are subjected to loss or 
gain such that the power or a modified form of it named as pseudo-power is conserved\cite{piju}. 
Soliton solution was found in the discrete non-linear Schr$\ddot{o}$dinger equation(DNLS) with 
gain and loss\cite{Suchkov}. The NLSE with balanced loss and gain has been investigated over 
the last  one decade or so and  is known to admit bright\cite{Bludov1,Driben1} and 
dark solitons \cite{Bludov}, breathers\cite{Barashenkov,Driben2}, rogue waves\cite{Kharif}, 
exceptional points\cite{Driben2,Konotop1}, etc. Further, the power oscillation is observed 
in some specific models which may have interesting technological applications\cite{pkg}. The 
NLSE with time-modulated loss-gain terms have also been investigated and shown its efficiency in 
the stabilization of solitons\cite{Driben2} and soliton switching\cite{Abdullaev3,Abdullaev4,Konotop1}. 
The NLSE with LC and BLG has also been investigated from the viewpoint of exact solvability and 
exact analytical solutions of some models are constructed under specific reduction of the original 
equation\cite{Driben1,Alexeeva}. Recently, a method was prescribed to construct exactly solvable 
models of NLSE with LC and BLG by using a non-unitary transformation which removes the LC and 
BLG terms completely at the cost of modifying the strength of the nonlinear term\cite{pkg}. 
Further, the LC and BLG terms are removed completely  without modifying the nonlinear
term at all for some specific models for which the non-unitary transformation can be identified 
as a pseudo-unitary transformation. The method has been used to construct exactly solvable 
models exhibiting power-oscillation. 

In this article, we consider a class of NLSE with time dependent LC and BLG as well as space-time 
modulated nonlinear term in an external complex potential. Both time dependent\cite{Abdullaev4} 
and constant\cite{Driben1,Driben2} BLG with particular types of NLSE and vanishing confining 
potential have been studied earlier\cite{Konotop1}. However, most of those studies are based 
on approximation and/or numerical methods . The sole purpose of this article is to  investigate 
the exact solvability of a generic system for various combinations of the time-modulated LC and 
BLG terms, space-time modulated nonlinear strengths and external potentials.

We follow a two-step approach. The LC and the BLG terms are removed completely in the first step 
via a non-unitary transformation by generalizing the technique prescribed in Ref. \cite{pkg}. 
This results in modifying the time-dependence of the nonlinear term and in general, a real-valued 
non-linear term becomes complex after the non-unitary transformation. We find that the LC and 
BLG terms are removed completely provided they have the same time-modulation. The method of 
separation of variables requires that the modified non-linear term is real. This is ensured 
by fixing the space-time modulated strengths of the nonlinear term. In the second step, the method 
of similarity transformation is used to investigate the solvability of the resulting equations. 
The method involves defining the differential equation in a new co-ordinate system accompanied 
by the multiplication of a scale-factor to the amplitude. This results in a system of 
equations with solvable limits for appropriate choice of co-ordinates and scale-factor.

We construct several exactly solvable models to exemplify the general method. The scheme 
for obtaining solvable limits for real and complex external potentials are different. 
Exact localized nonlinear modes with spatially constant phase may be obtained for a real 
potential for which the corresponding linear Schr$\ddot{o}$dinger equation is solvable. 
Exact analytical solutions with space-time dependent phase are also obtained for a few real 
potential. The complex potential offers more flexibility in comparison with real potential in 
constructing solvable models. We devise a general method based on supersymmetric quantum 
mechanics\cite{khare} to find a large class of complex potentials admitting localized  nonlinear 
modes. In general, phases of the nonlinear modes are space-time dependent. 
Phase singularity is present in the nonlinear mode for a few systems with
complex potentials.  

A few model independent features of the exact solutions are the following. The exact solutions are
insensitive to the detail form of the space-modulated nonlinear strength. It is not obvious whether
this is specific to the ansatz chosen for obtaining exact solutions or a manifestation of some underlying
symmetry. Further, the power-oscillation in time is absent for time-independent non-linear strength. There are
NLSE with constant LC, BLG and nonlinear strength for which power-oscillation has been observed\cite{pkg}.
The class of NLSE considered in this article has nonlinear interaction which is different from the model
considered in Ref. \cite{pkg}. Thus, the power oscillation in time seems to be dependent on the nonlinear
interaction.

The plan of the article is the following. The model is introduced in Sec. II along with Lagrangian and Hamiltonian
formulations of the system. In Sec. III, the two-step approach is implemented to map the original equation into
a solvable model. Explicit examples are constructed for systems without and with external real potentials in Sec. 
IV and V, respectively. Results pertaining to complex confining potential are presented in Sec. VI. Finally,
the findings are summarized in Sec. VII with discussions and outlook.
In Appendix-I, solutions of a nonlinear equation arising in the discussions of solvable models are described.

\section{The Model}

We consider one dimensional 2-component NLSE as
\bea
i\psi_{1t} & = & - \psi_{1xx} + k^{*}(t) \psi_{2} + i \gamma(t) \psi_{1} + V(x) \psi_{1} 
+ \left(g_{11}(x,t) |\psi_1|^{2} + g_{12}(x,t) |\psi_{2}|^{2}\right) \psi_1 \nonumber \\
i\psi_{2t} & = & - \psi_{2xx} + k(t) \psi_{1} - i \gamma(t) \psi_{2} + V(x)\psi_{2} 
+ \left(g_{21}(x,t) |\psi_1|^{2} + g_{22}(x,t) |\psi_{2}|^{2}\right) \psi_2
\label{nlse1}
\eea
\noindent where $k(t)$ is the strength of the linear coupling among the two fields and $k^{*}$ denotes
the complex conjugate of $k$. The time-modulated strength of the balanced loss gain terms is $\gamma(t)$.
The space-time modulated strengths of the non-linear interaction are denoted as $g_{ij}(x,t)$.
In the terminology of optics,  $g_{11}, g_{22}$ are self phase modulation terms, while $g_{21},g_{12}$
correspond to cross-phase modulation. The external potential $V(x)$ is relevant
for the mean-field description of Bose-Einstein condensates. Several ${\cal{PT}}$-symmetric complex potentials
$V(x)$ exhibit many interesting phenomenon. We discuss both real and complex potentials separately in this article
and choose $V(x)=s(x)+i\tilde{s}(x)$, where functions $s(x), \tilde{s}(x)$ are real.
There are several solvable limits of Eq.(\ref{nlse1}), starting from the simplest case of
$g_{ij} = 0, k = \gamma =0$ which models the optical wave propagation under paraxial approximation\cite{Ganainy,Markis}.
The integrable canonical form of Manakaov-Zakharov-Schulman(MZS)
system\cite{kk,zs} is obtained for constant coefficients $g_{ij}$ with $g_{11}=g_{21}, g_{12}=g_{22}$ and vanishing
linear coupling, loss-gain terms and external potential.
Exact solutions of MZS system with non-vanishing constant loss-gain terms and time-dependent $g_{ij}$ with specific 
time-modulation have been constructed via non-unitary transformations\cite{pkg}. The solutions for constant $g_{ij}$ appear
as a special case. The Gross-Pitaevskii equation, which provides a mean-field description of BEC, is obtained for vanishing
loss-gain terms and non-vanishing $V(x)$. Several exact solutions for different choices of $V(x)$ and space-modulated
nonlinear strengths have been constructed\cite{Beitia}. It appears that solvable limits of Eq. (\ref{nlse1}) in its generic form,
particularly with time-dependent $k, \gamma$, space-time dependent $g_{ij}$ and with or without the external potential $V(x)$,
have not been investigated so far. The method of non-unitary transformation\cite{pkg} will be employed to investigate
solvability of Eq. (\ref{nlse1}) in its generic form.

It is convenient to rewrite Eq.(\ref{nlse1}) in a compact form in terms of the Pauli matrices $\sigma_1, \sigma_2,
\sigma_3$, $2 \times 2$ identity matrix $\sigma_0$ and two-component complex vector $\Psi(x,t)=(\psi_1,\psi_2)^T$,
where the superscript $^T$ denotes transpose. We define projection operators $P_{\pm}=\frac{1}{2}\left ( \sigma_0
\pm \sigma_3 \right )$ with the properties $P_+ \Psi=(\psi_1, 0)^T, P_- \Psi=(0,\psi_2)^T$ and the matrices
$\sigma_{\pm}:= \frac{1}{2} \left (\sigma_1 \pm i \sigma_2 \right )$. The matrices $P_{\pm}, \sigma_{\pm}$ form
a basis for $2 \times 2$ matrices. We introduce a non-hermitian scalar gauge
potential $A(t)$ and an operator $D_0$,
\bea
A(t) = k \sigma_- + k^* \sigma_+ + i \gamma \sigma_{3}, 
D_0:= \sigma_0 \frac{\partial}{\partial t} + i A(t).
\label{adexp}
\eea
\noindent The term $i \gamma(t) \sigma_3 \Psi$ corresponds to time-dependent BLG, while
$ k(t) \sigma_- + k^*(t) \sigma_+$ denote time-dependent LC. The time-dependence of $A(t)$ can
be controlled by choosing $k(t)$ and $\gamma(t)$ appropriately. The operator $D_0$ resembles the temporal
component of covariant derivative with non-hermitian scalar gauge-potential $A$. 
It may be noted that we have not included any term of the form $f(t) \sigma_0$ to $A$, since it can always be trivially
gauged away through a $U(1)$ transformation of the form $\Psi \rightarrow e^{-i \int^t f(t^{\prime}) dt^{\prime} } \Psi$.
The $U(1)$ phase-factor $e^{-i \int^t f(t^{\prime}) dt^{\prime} }$ does not affect the dynamical variables
like power, width of the wave-packet and its speed. Thus, the form of $A$ is quite general.

Eq. (\ref{nlse1}) can be rewritten as,
\bea
i D_0 \Psi & = & -  \Psi_{xx} + V(x) \Psi
 + \left [ \left ( g_{11}(x,t)  \Psi^{\dagger} P_{+} \Psi + g_{12}(x,t) \Psi^{\dagger} P_{-} \Psi \right ) P_{+} \right. \nonumber \\
           & + & \left. \left ( g_{21}(x,t) \Psi^{\dagger} P_{+} \Psi + g_{22}(x,t) \Psi^{\dagger} P_{-} \Psi \right ) P_{-} \right ] \Psi 
\label{nlse2}
\eea
\noindent which is convenient for further analysis using non-unitary transformation. Defining the power $P=\Psi^{\dagger} \Psi$ and
$Q=\int dx P$, it can be checked easily that $Q$ is not a constant of motion for $\gamma \neq 0$. The power-oscillation is a hallmark
of some systems with balanced loss-gain and it is a manifestation of fact that $Q$ is not a constant of motion. In fact,
the expression of $P$ in Ref. \cite{pkg} contains a time-dependent periodic-function which becomes unity in the limit of vanishing
loss-gain terms multiplied by a space-dependent function. The quantity $\tilde{Q}=\int dx \tilde{P}$ with
$\tilde{P}=\Psi^{\dagger} \eta \Psi$ is a constant
of motion for $\gamma \neq 0, g_{11}=g_{21}, g_{12}=g_{22}$, provided the matrix $A$ is $\eta$-pseudo-hermitian, i.e.
$A^{\dagger}=\eta A \eta^{-1}$. The positive-definite matrix $\eta$ for $A$ in Eq. (\ref{adexp}) is given in Ref. \cite{pkg,piju}.
The system admits a Lagrangian and Hamiltonian formulation
for  $g_{12} = g_{21} $:
\bea
{\cal{L}} & = & \frac{i}{2} \left [ \Psi^{\dagger} D_0 \Psi - \left ( D_0 \Psi \right )^{\dagger} \Psi \right ]  +
\frac{1}{2} \Psi^{\dagger} \left (A^{\dagger} - A \right ) \Psi - {\Psi_{x}}^{\dagger} \Psi_{x} 
- V(x) \Psi^{\dagger} \Psi  -  \frac{g_{11}}{2} \left ( \Psi^{\dagger} P_{+} \Psi \right )^{2} \nonumber \\
	  & - & \frac{g_{22}}{2} \left ( \Psi^{\dagger} P_{-} \Psi \right )^{2} 
 - g_{12} \left ( \Psi^{\dagger} P_{+} \Psi \right ) \left ( \Psi^{\dagger} P_{-} \Psi \right ),\nonumber \\
{\cal{H}} & = & \Psi_x^{\dagger} \psi_x + V(x) \Psi^{\dagger} \Psi + \Psi^{\dagger}
A \Psi + \frac{g_{11}}{2} \left ( \Psi^{\dagger} P_{+} \Psi \right )^{2} 
+ \frac{g_{22}}{2} \left ( \Psi^{\dagger} P_{-} \Psi \right )^{2}
+ g_{12} \left ( \Psi^{\dagger} P_{+} \Psi \right ) \left ( \Psi^{\dagger} P_{-} \Psi \right ).
\label{lang-hami}
\eea
\noindent The Hamiltonian density ${\cal{H}}$ is complex-valued for $\gamma \neq 0$ and the corresponding quantum
${\cal{H}}$ is non-hermitian with appropriate quantization condition. For $g_{11}=g_{22}=g_{12}$, the Lagrangian density
is invariant under global phase transformation $\Psi \rightarrow e^{i \theta \eta} \Psi, \theta \in \mathbb{R}$
and the corresponding conserved charge is $\tilde{Q}$.

\section{Transformation to solvable equation}

The purpose of this section is to generalize the technique outlined in Ref. \cite{pkg}
to remove the time-modulated BLG and LC by using a non-unitary transformation. The transformation
modifies the non-linear term and suitable choices of the space-time modulated coefficients
lead to solvable models. We consider a transformation relating $\Psi$ to a two-component
complex scalar field $\Phi$ as follows,
\bea
\Psi(x,t) = U(t) \Phi(x,t) 
\label{non-uni}
\eea
\noindent It will appear later that the operator $U(t)$ in general is non-unitary.
The BLG and LC terms are contained only in $D_0 \Psi$ and it can be checked
that $D_0 \Psi = U \partial_t \Phi$
provided $U$ satisfies the equation,
\bea
\frac{dU}{dt} = -i A U(t)
\label{condi-a}
\eea
\noindent A similar equation appears in the study of scattering theory in the interaction picture
and the operator analogous to $U$ in that context is known as Dyson operator,
which describes the time-evolution of a state $|\Psi(t_0)\rangle$ to $|\Psi(t)\rangle$, i.e.
$|\Psi(t)\rangle=U(t,t_0) |\Psi(t_0)\rangle$. The operator $U$ can not be identified as
Dyson operator in the present case since it maps $\Phi(x,t)$ to $\Psi(x,t)$ at the same time $t$ and
the physical context is also different. However, the solution of $U$ in Eq. (\ref{condi-a}) may be
obtained as an infinite series along the line of derivation of Dyson series. We are interested
in a closed form expression of $U(t)$ from the viewpoint of exact solvability. This leads to the
solution of $U(t)$ for specific choice of $A$ as,
\bea
U(t) = e^{-i \int^t A(t^{\prime}) dt^{\prime}},\ \
\left [ A, \int^{t} A(t') dt' \right ]=0
\label{condi-b}
\eea
\noindent Note that $U(t)$ is non-unitary, since $A(t)$ is non-hermitian.
Unitary transformations have been used in physics in different contexts, particularly
in the context of field theory, for past several decades. To the best of our knowledge,
the use of non-unitary transformation to construct exact solution in a systematic way
has not been considered earlier. Within this background, it may be noted that a unitary
transformation is a change of basis in the field-space by keeping the norm
fixed, while the norm is not preserved under a non-unitary transformation. This is
a major difference \textemdash systems related by unitary transformation are gauge equivalent,
while the same can not be claimed for systems related by non-unitary transformation.
This is manifested in the result that the power of the standard Manakov system is dif-
ferent from the Manakov system with balanced loss-gain, although they are connected
via a non-unitary/pseudo-unitary transformation\cite{pkg}. The same is true for the system
considered in this article, since $\Psi^{\dagger}\Psi \neq \Phi^{\dagger}\Phi$. Similarly,
one can show that the time-dependence of other observables like width of the wave-packet and its speed
of growth are different for systems connected via non-unitary/pseudo-unitary trans-
formation. However, for systems connected by unitary transformation, observables
like power, width of the wave-packet and its growth are identical. The unitary trans-
formation only mixes different components of the field leading to different expression
for the solutions.

The second condition of (\ref{condi-b}) implies that $k(t)$ and $\gamma(t)$ should have the same
time dependence.  We choose
$k(t) = \mu_0(t) \beta, \gamma(t)=\mu_0(t) \Gamma$ in terms of an arbitrary real function $\mu_0(t)$ and
$\beta \in \mathbb{C}, \Gamma \in \mathbb{R}$ leading to the following expression of $A(t)$:
\bea
A(t)=\mu_0(t) A_0, \ A_0 := \beta \sigma_- + \beta^* \sigma_+ + i \Gamma \sigma_3.
\eea
\noindent We introduce the following quantities:
\bea 
\mu(t)=\int^t \mu_0(t^{\prime}) dt^{\prime}, \ \epsilon_0= \sqrt{{\vert \beta \vert}^2 - \Gamma^2}, \
\epsilon(t) = \mu(t) \epsilon_0.
\label{mu1}
\eea
\noindent The qualitative behaviour of the operator $U(t)$ primarily depends on $\epsilon_0$
being positive, zero and purely imaginary. We denote the corresponding $U(t)$ as $U_{+}(t), U_0(t)$ and
$U_I(t)$, respectively, with the  following expressions:
\bea
U_+(t) & = & \sigma_{0} \cos(\epsilon) - \frac{i {A}_0}{\epsilon_0}  \sin(\epsilon), \epsilon_0 > 0\nonumber \\
U_0(t) & = & \sigma_0 - i A_0 \mu(t)\nonumber \\
U_I(t) & = & \sigma_{0} \cosh({\vert \epsilon_0 \vert} \mu(t)) -
i \frac{A_0}{\vert \epsilon_0 \vert} \sinh({\vert \epsilon_0 \vert} \mu(t)) 
\label{U} 
\eea
\noindent The effect of the time-dependent scale-factor $\mu_0(t)$ in $A$ is to change the time-modulation
of $U(t)$. For a time-independent $\mu_0(t)$, i.e. $\epsilon(t)=\epsilon_0 t$, the matrix $U_+(t)$ becomes periodic
in time. However, the matrices $U_0(t)$ and $U_I(t)$ becomes unbounded and the solutions
$\Psi(x,t)$ becomes unbounded even for a bounded $\Phi(x,t)$. An interesting point to note is that
bounded sloutions may be obtained for all three cases, namely, $U_+, U_0, U_I$, by suitably choosing
the $\mu_0(t)$. For example, $\mu(t)$ reduces to the Error function, i.e. $\mu(t)=erf(t)$, for the choice
\bea
\mu_0(t)=\frac{2}{\sqrt{\pi}} e^{-t^2} 
\eea
\noindent and $U(t)$ corresponding to all the three cases discussed above are bounded.
Thus, appropriate time-modulation may be used to stabilize a system whose unboundedness
comes solely from $U(t)$.

The removal of the BLG and LC terms through the non-unitary transformation modifies the
nonlinear term and imparts additional time-dependence on it. In particular, substituting the
expression of $\Psi(x,t)$ in Eq. (\ref{non-uni}) into Eq. (\ref{nlse2}), yields the following
\bea 
i \Phi_{t}  & = & - \Phi_{xx} + V(x) \Phi + \left [  
K(g_{11}, g_{21}) \ \Phi^{\dagger} F_+ \Phi 
+ K(g_{12}, g_{22}) \ \Phi^{\dagger} F_- \Phi \right ] \Phi, 
\label{t-nlse1}
\eea
\noindent where $ K(\xi_1, \xi_2) = \xi_1 \left ( U^{-1} P_{+} U \right ) + \xi_2 \left ( U^{-1} P_{-} U \right )$ and
$F_{\pm} = U^{\dagger} P_{\pm} U$.
With the introduction of the functions
$T_{\pm}$,
\bea
T_{\pm} = \frac{\Gamma \mu(t)}{\epsilon^2} \sin^2(\epsilon) \pm \frac{\sin(2\epsilon)}{2\epsilon},
\eea
\noindent the explicit expression of $F_{\pm}$ and $K$ may be obtained as follows:
\bea
F_{\pm}        & = &  \left (\frac{1}{2}+\Gamma \mu(t) T_{\pm} \right ) \sigma_0
-i T_{\pm} \mu(t) \left ( \beta^* \sigma_+ - \beta \sigma_- \right )
 + \left ( \frac{\Gamma \mu(t)}{2\epsilon} \sin(2\epsilon) \pm \frac{1}{2}\cos(2\epsilon) \right )
\sigma_{3},\nonumber \\ 
K(\xi_1,\xi_2) & = & \frac{\xi_1+\xi_2}{2} \sigma_0 + i \left ( \xi_1 -\xi_2\right ) \mu(t) 
\left ( \beta^* \sigma_+ T_- + \beta \sigma_- T_+ \right )
 + \left ( \xi_1-\xi_2 \right )  
\left (\frac{1}{2}  - \frac{\mu^2(t){\vert \beta \vert}^{2}}{\epsilon^2} \sin^2(\epsilon) \right ) \sigma_3 
\label{K}
\eea
\noindent We have presented the above expressions for the generic allowed values of $\epsilon_0$ and appropriate
limits may be taken depending on the physical context of the problem.
It may be noted that $T_{\pm}$ is always real valued independent of whether $\epsilon_0$ is positive, zero
or purely imaginary. Further, the operator $F_{\pm}$ is necessarily hermitian, while $K(\xi_1, \xi_2)$ is 
hermitian either for (i) $\xi_1 = \xi_2$ or (ii) $\xi_1 \neq \xi_2, T_+ = -T_-$. The condition $T_+=-T_-$ 
is achieved for $\Gamma \mu(t)=0$. The choice $\mu(t)=0 \ \forall \ t$ is excluded, since LC as well as 
BLG terms also vanish in this limit. Thus, for $\xi_1 \neq \xi_2$, $K(\xi_1, \xi_2)$ is hermitian 
for $\Gamma = 0$. i.e. the limit of vanishing loss-gain terms. In general, the nonlinear term in 
Eq. (\ref{t-nlse1}) is non-hermitian \textemdash the non-unitary transformation involving $U(t)$ removes the 
loss-gain terms at the cost of introducing
non-hermiticity in the non-linear term.  The nonlinear term in Eq. (\ref{t-nlse1}) becomes hermitian
either for (i) $ g_{11}=g_{21}, g_{22}=g_{12}, \Gamma \neq 0$ or (ii)$\Gamma = 0$,
since both $K(g_{11}, g_{21})$ and $K(g_{11}, g_{21})$ are hermitian in these limits. The second condition
corresponds to vanishing loss-gain term for which $U(t)$ is unitary and the result is expected.
The nonlinear term for the first condition takes the form of MZS system\cite{kk,zs}
$\left [ \Psi^{\dagger} \left ( g_{11} P_+ + g_{22} P_- \right ) \Psi \right ] \Psi$, which remains
real-valued after the non-unitary transformation. This particular form of nonlinear interaction
with $V(x)=0, \mu_0(t)=1$ and space-time independent $g_{11}, g_{22}$ has been investigated earlier\cite{pkg}.

A pertinent question at this juncture is whether or not the solution $\Psi$ of Eq. (\ref{nlse2})
corresponding to a stable solution $\Phi$ of Eq. (\ref{t-nlse1}) is also stable. The answer is that
the transformation (\ref{non-uni}) that connects the original system described by Eq. (\ref{nlse2}) to the
Eq. (\ref{t-nlse1}) does not alter the stability property of $\Phi$ for a bounded $U(t)$. In order
to see this, we consider $\Phi(x,t)= \tilde{\Phi}(x,t) + \chi(x,t)$, where $\tilde{\Phi}(x,t)$
is an exact solution of Eq. (\ref{t-nlse1}) and  $\chi(x,t)$ is small perturbation. Plugging the
expression of $\Phi(x,t)$ in Eq. (\ref{t-nlse1}), we obtain the following equation in the leading
order of the perturbation,
\bea
i \chi_{t}  & = & - \chi_{xx} + V(x) \chi + \left [ K(g_{11},g_{21}) \left\{ \tilde{\Phi}^{\dagger} F_{+}
\chi + \chi^{\dagger}F_{+}\tilde{\Phi} \right\} + K(g_{12},g_{22})
\left \{ \tilde{\Phi}^{\dagger}F_{-}\chi \right. \right. \nonumber \\
            & + & \left. \left. \chi^{\dagger}F_{-}\tilde{\Phi} \right\}\right] \tilde{\Phi}
            + \left[ K(g_{11},g_{21}) \tilde{\Phi}^{\dagger}F_{+}\tilde{\Phi} +
K(g_{12},g_{22}) \tilde{\Phi}^{\dagger}
            F_{-} \tilde{\Phi} \right] \chi + {\cal{O}} \left (\chi^2 \right )
\label{stab2}
\eea
\noindent The Eq. (\ref{stab2}) determines the stability of the exact solution $\tilde{\Phi}$  and
the solution is stable if $\chi$ is a bound state. The exact solution of Eq,  (\ref{nlse2}) corrresponding
to $\tilde{\Phi}$ is $\tilde{\Psi}=U \tilde{\Phi}$. We consider $\Psi=\tilde{\Psi} + U \eta(x,t)$, where 
$U(t)$ is taken to be bounded in time and $\eta(x,t)$ is an arbitrary function. The perturbation to
the exact solution $\tilde{\Psi}$ is $U \eta(x,t)$ and arbitrariness of the small fluctuations is
contained in $\eta(x,t)$. We find after substituting the expression of $\Psi(x,t)$ in Eq. (\ref{nlse2})
and keeping only the leading order terms that $\eta$ satisfies the Eq. (\ref{stab2}). Thus, the stability
of the original Eq. (\ref{nlse2}) and the transformed Eq. (\ref{t-nlse1}) is governed by the same Eq.
(\ref{stab2}) and the transformation (\ref{non-uni}) can not change the stability property for a bounded
$U$ in time and under identical initial conditions.

The investigations on the complete integrability of the system described by Eq. (\ref{t-nlse1}) is a highly nontrivial
problem in presence of $V (x)$, space-time modulated nonlinear-strengths and loss-gain terms. We
are interested in this article in finding solvable limits of Eq. (\ref{nlse1}) so that the exact 
solutions can be used in plethora of physical systems in which it appears. The reduction of 
Eq. (\ref{nlse1}) to Eq. (\ref{t-nlse1}) with a closed form
expression for $U(t)$ has been performed on general ground. In order to proceed further
for analyzing two-component vector non-linear Schr$\ddot{o}$dinger equation of the form given
in Eq. (\ref{t-nlse1}), we use the method of separation of variables by choosing
one of the standard ansatzes which is consistent with the group-theory based analysis\cite{Beitia} of Eq. (\ref{t-nlse1})
for time-independent and real-valued nonlinear strength. We consider the solution of  Eq. (\ref{t-nlse1}) as,
\begin{equation}
\Phi(x,t) = W \ R(x) \ e^{i\big(\theta(x) - Et\big)},
\label{ansatz}
\end{equation}
\noindent where $W=( W_1 e^{i \theta_{1}}, W_2 e^{i \theta_{2}} )^T$ is a two-component constant, complex vector,
$R(x)$ and $\theta(x)$ are real functions of
their arguments and $E$ is a constant. Inserting the above expression into Eq. (\ref{t-nlse1}) we find,
\bea
R_{xx} & = & -i\left (R \theta_{xx} + 2 R_{x} \theta_{x} - R \tilde{s}(x) \right ) - \left (E - s(x) \right ) R + 
R \theta^{2}_{x} \nonumber \\
       & + & \frac{1}{W^{\dagger} W} \left [ W^{\dagger} K(g_{11}, g_{21}) W \ W^{\dagger} F_+ W
 +  W^{\dagger} K(g_{12}, g_{22}) W \ W^{\dagger} F_- W \right ] R^3(x).
\label{R-eqn}
\eea
\noindent It is to be noted that in general the potential $V(x)$ in Eq.(\ref{t-nlse1}) is complex and has 
the form $V(x) = s(x) + \tilde{s}(x)$. The nonlinear term is complex and time-dependent. The method of 
separation of variables with the ansatz as above fails unless the coefficient of the nonlinear term is 
real and time-independent, which can be achieved with the judicious choice of the space-time modulated 
coefficients $g_{ij}(x,t)$. The imaginary part of the coefficient vanishes for the condition,
\bea
\frac{g_{11}-g_{21}}{g_{12}-g_{22}} = -\frac{W^{\dagger} F_- W}{W^{\dagger} F_+ W}
\label{condi-g}
\eea
\noindent while the real part of the coefficient of the nonlinear term is time-independent and equals to
$f(x)=\frac{1}{2} \left ( f_1(x) + f_2(x) \right )$ provided,
\bea
g_{11}+g_{21} =\frac{f_1(x)}{W^{\dagger} F_+ W}, \ \ g_{12} + g_{22} = \frac{f_2(x)}{W^{\dagger} F_- W}
\label{geqn-2}
\eea
\noindent where $f_1(x)$ and $f_2(x)$ are arbitrary functions.
The space-time modulated strength $g_{ij}$ of the nonlinear interaction may be obtained by solving Eqs. (\ref{condi-g})
and (\ref{geqn-2}) which constitute an undetermined system. We solve the equations by keeping $g_{22}(x,t)$ arbitrary:
\bea
g_{11}(x,t) & = & \frac{1}{W^{\dagger} F_+ W} \left [ f_1(x) - f(x) + g_{22}(x,t) W^{\dagger} F_- W \right ],\nonumber \\
g_{21}(x.t) & = & \frac{1}{W^{\dagger} F_+ W} \left [ f(x) - g_{22}(x,t) W^{\dagger} F_- W\right ],\nonumber \\ 
g_{12}(x.t) & = & \frac{1}{W^{\dagger} F_- W} \left [ f_2(x) - g_{22}(x,t) W^{\dagger} F_- W \right ]
\label{g-general}
\eea
\noindent It should be emphasized that each choice of $g_{22}(x,t), f_1(x), f_2(x), \mu(t)$ leads to a
different classes of the space-time modulation $g_{ij}$ of the nonlinear term. However, the nonlinear equation
(\ref{R-eqn}) depends only on $f(x)$ and does not keep track of the specific forms of $g_{22}(x,t), f_1(x), f_2(x)$,
and hence of $g_{ij}$, for fixed $f(x)$. This implies that the spatial dependence of the solution of Eqn. (\ref{nlse1})
in terms of $R(x)$ and $\theta(x)$ is same for a large class of models characterized by different
$g_{ij}$. This result is very important, since solvability of Eq. (\ref{R-eqn}) leads to solvability
of a very large class of NLSE with LC and BLG terms characterized by various forms of $g_{ij}$.

We choose a symmetric form of the $g_{ij}$ for presenting our results:
\bea
&& g_{11} = \frac{ f_1(x) + G(x,t) }{2 W^{\dagger} F_+ W},
g_{21} = \frac{ f_1(x) - G(x,t) }{2 W^{\dagger} F_+ W}, \nonumber \\
&& g_{22} = \frac{ f_2(x) + G(x,t) }{2 W^{\dagger} F_- W},
g_{12} = \frac{ f_2(x) - G(x,t) }{2 W^{\dagger} F_- W},
\label{geqn-3}
\eea 
\noindent where $G(x,t)$ is an arbitrary function. The introduction of $G(x,t)$ is to keep track of the
arbitrariness present in the solution of the undetermined system of Eqs. (\ref{condi-g}) and (\ref{geqn-2}) so that
the generality is not lost. The expressions for $g_{11}, g_{21}$ and $g_{22}$ may be obtained by substituting
$g_{22}$ from Eq. (\ref{geqn-3}) in Eq. (\ref{g-general}).  The expressions for $W^{\dagger} F_{\pm} W$ are,
\bea
W^{\dagger} F_{\pm} W & = & b_0 \left ( \frac{1}{2} + T_{\pm}(t) \mu(t) D \right )
 + b_3 \left [ \frac{\Gamma \mu(t)}{2\epsilon} \sin(2 \epsilon) \pm \frac{1}{2} \cos (2 \epsilon) \right ]
\label{wpmw}
\eea
\noindent where $\beta = {\vert \beta \vert} e^{i \theta_{3}}$, $b_j=W^{\dagger} \sigma_j W,
j=0, 1, 2, 3$ and the constant $D$ is defined as,
\bea
D \equiv \Gamma + \frac{2}{b_0} W_1 W_2 {\vert \beta \vert} \sin \left (\theta_{2}- \theta_{1} -\theta_{3} \right ).
\label{D-eqn}
\eea
\noindent  The time-dependence of the space-time 
modulated coefficients is related to the choice of $\mu(t)$ and $G(x,t)$. There are
several interesting possibilities, including time-independent $W^{\dagger} F_{\pm} W=\frac{b_0}{2}$ for
$b_3=D=0$. The choice $W_1=W_2\equiv W$ leads to $b_3=0$ and $D=0$ gives the relation
$\Gamma = - {\vert \beta \vert} \sin \left (\theta_{2}- \theta_{1} -\theta_{3}
\right ) $. The condition $\Gamma^2 < {\vert \beta \vert}^2$ for time-periodic $U(t)$ is ensured provided
$\theta_2-\theta_1-\theta_3 \neq (2n+1) \frac{\pi}{2}, \ n \in \mathbb{Z}$. Thus, $g_{ij}$'s are also independent
of time in this limit provided $G(x,t)$ is taken as time independent.
The space-dependence may be tailored by choosing appropriate functions $f_1(x), f_2(x), G(x,t) \equiv G(x)$. The choice
$G(x,t)=0$ corresponds to $ g_{11}=g_{21}, g_{22}=g_{12}$ which has been noted earlier as the limit for hermitian $K$.

The imaginary part of the nonlinear interaction of  Eq. (\ref{R-eqn}) vanishes for $g_{ij}$'s chosen as
in Eq.(\ref{geqn-3}). The imaginary and real parts of Eq. (\ref{R-eqn}) can be  separated as,
\bea
\label{theta0}
&& \Big(R^{2} \theta_{x}\Big)_{x} = \tilde{s}(x) R^{2},\\
&& R_{xx} + \big(E - s(x)\big) R - R {\theta_{x}}^{2} = f(x) R^{3}.
\label{R1}
\eea
\noindent Note that the above equations are independent of $G(x,t)$, although the space-time modulated coefficients
$g_{ij}$ explicitly depend on it. This is a surprising result that $G(x,t)$ has no effect at all on the solutions
$\Psi$. It seems that the specific ansatz for $\Phi$ in Eq. (\ref{ansatz}) leads to this result. The system defined
by Eq. (\ref{nlse1}) with $g_{ij}$ given by Eq. (\ref{geqn-3}) may admit solutions which depend on the choice 
of $G(x,t)$. However, the chosen ansatz is not suitable for any such exploration \textemdash different
analytic and/or numerical methods may have to be employed for the purpose which is beyond the scope of
this article. The solutions of Eqs. (\ref{theta0},\ref{R1}) do not depend on $f_1(x)$ and $f_2(x)$ either, but
on their average $f(x)$. This is again possibly related to the chosen ansatz and has to be confirmed
independently through other means. The task is to solve Eqs. (\ref{theta0},\ref{R1}) for given $s(x), \tilde{s}(x)$ and $f(x)$
which characterize the forms of the potential and space-modulation of the nonlinear strengths, respectively. 
The method involves defining a new co-ordinate $\zeta(x)$ and expressing
$R(x)$ as the product of a scale-factor $\rho(x)$ and $\zeta$-dependent function $u(\zeta)$,
\bea
R(x) = \rho(x) \ u \big(\zeta(x)\big), \ \zeta(x) = \int^{x} \frac{ds}{\rho^2(s)}.
\label{zeta}
\eea
\noindent The treatment for obtaining exact solutions for real and complex potentials are different and discussed
separately:\\

{\bf Real Potential}: The imaginary part vanishes, i.e. $\tilde{s}(x) = 0$ and $\theta(x)$ is determined from
Eq.(\ref{theta0}) as,
\bea
\theta(x) = \int \frac{C}{R^{2}} dx,
\label{theta}
\eea 
\noindent where $C$ is an integration constant. The decoupled equation for $R(x)$ is obtained by substituting
$\theta(x)$ into Eq.(\ref{R1}): 
\bea
R_{xx} + \big(E - s(x)\big) R - \frac{C^2}{R^3} = f(x) R^{3}.
\label{R3}
\eea
\noindent Eq. (\ref{R3}) reduces to the famous Ermakov-Pinney equation\cite{ep} for $f(x)=0$, while $C=0=s(x), f(x)=1$ leads
to the standard NLSE. The equation (\ref{R3}) is solvable in both the limits. There is a very interesting reduction of
Eq. (\ref{R3}) for $C=0=f(x)$ for which it reduces to the linear Schr$\ddot{o}$dinger equation.
The vanishing constant $C$ implies that the phase $\theta$ is constant. The condition $f(x)=0$ can be implemented by
taking $f_2(x)=-f_1(x)$, where $f_1, f_2$ may be considered as constants or space-dependent. Eq. (\ref{R3}) reduces
to the standard linear Schr$\ddot{o}$dinger equation with real potential $s(x)$ for $C=0=f(x)$. We have the important
result that the system admits exact localized nonlinear modes for any $s(x)$ for which the corresponding linear
Schr$\ddot{o}$dinger equation is solvable. We do not present any example in this regard.
For the general case, the substitution of Eq.(\ref{zeta}) into Eq.(\ref{R3}) results in the
following sets of equations:
\bea
\label{u1}
&& u_{\zeta\zeta} + m u - \frac{C^2}{u^3} - 2 \sigma u^3 = 0, \\
&& \rho_{xx} + \big(E - s(x)\big) \rho  = \frac{m}{\rho^3}, \ \
f(x) = \frac{2\sigma}{\rho^6},
\label{u1.5}
\eea   
\noindent where $m$ and $\sigma$ are real constants. The $\theta$ dependence is contained solely in Eq. (\ref{u1}).
The solutions of Eq. (\ref{u1.5}) for a given $s(x)$ is used to fix $f(x)$ and $\xi(x)$. The equation for $u$
can be solved independently and substitutions of $\xi(x)$ along with $\rho(x)$ determines $R(x)$. Exact solutions
of Eq. (\ref{u1}) is discussed in Appendix-I.

Note that Eq. (\ref{u1.5}) reduces to the linear Schr{\"o}dinger equation with $\rho$ playing the role
of the eigenfunction for $m=0$. The complete spectra  of the linear equation are known
for a large number of potential $s(x)$. Each eigenstate $\rho$ for a given $s(x)$  and $E$,
determines  $f(x)=\frac{2 \sigma}{\rho^6}$ trivially, which corresponds to a unique Eq. (\ref{nlse2})
via the $f(x)$ dependence of $g_{ij}$. It should be noted that different $\rho$ corresponding to different
$E$ for a given $s(x)$ does not correspond to linearly independent solutions of Eq. (\ref{nlse2}) for fixed $ g_{ij}$,
rather it defines different NLSE. Thus, the method can be used to find exact solution of a large class
of NLSE given by Eq. (\ref{nlse2}). If closed from expressions
for the integrations appearing in Eqs. (\ref{zeta}) and (\ref{theta}) are available, a complete analytic solution
for $\Psi$ may be obtained. This only shows that Eq. (\ref{nlse2}) with a wide class of nonlinear strengths
are amenable to exact solutions by using the method proposed herein. 
We shall present only a physically relevant
prototype for a given $V(x)$ and $f(x)(g_{ij})$ to exemplify the general method in Sec. V \& VI.

{\bf Complex Potential:} The function $\theta(x)$ can not be expressed as the integral of $R(x)$ alone 
due to non-vanishing imaginary part of the potential i.e. $\tilde{s}(x) \neq 0$.
Substitution of Eq.(\ref{zeta}) into Eq.(\ref{R1}) results in the following sets of equations
\bea
\label{u2}
&& u_{\zeta\zeta} + mu - 2\sigma u^3 = 0 \\
&& \frac{\rho_{xx}}{\rho} + E - {\theta_{x}}^{2} - \frac{m}{\rho^4} = s(x),
\ \ f(x) = \frac{2\sigma}{\rho^6},
\label{u2.5}
\eea
\noindent which are completely different from Eqs. (\ref{u1},\ref{u1.5}) for the case of real potential. The effect of the
imaginary part is contained in Eq. (\ref{u2.5}) via $\theta_x$ and taking the limit of vanishing $\tilde{s}(x)$ does
neither reproduce Eq. (\ref{u1.5}) nor a decoupled equation for $\rho$ is obtained. The scheme for constructing
the solvable system is the following \textemdash  we fix $g_{ij}$ by choosing constant $f(x)$, say $f(x)= 2 \sigma$ for
simplicity, which determines $\rho$, and a relation between $\theta_x$ and $s(x)$,
\bea
\theta_x^2=E-m-s(x) .
\label{tt}
\eea
\noindent The complex potential is chosen such that the Eqs.(\ref{theta0}, \ref{R1}) and (\ref{tt}) are 
consistent. This prescription is applicable for specific choices of $g_{ij}$ and complex potentials,
nevertheless, it exhausts a large class of exactly solvable models.

The solution of Eq. (\ref{nlse1}) has the form,
\bea
\Psi(x,t) = U(t) \ W \ \rho(x) \ u\big(\zeta(x)\big) \ e^{i(\theta(x) - Et)}, 
\eea
\noindent where the expressions for $\rho(x), u(\zeta)$ is determined from Eqs. (\ref{u1},\ref{u1.5}) for
the real potential and from Eqs. (\ref{u2},\ref{u2.5}) for the complex potential.  The power $P(x,t)$ is factorised  in terms of
time-dependent and space-dependent parts as,
\bea
P(x,t)=P_1(t) R^2(x), P_1(t) \equiv W^{\dagger} U^{\dagger}(t) U(t) W.
\label{exp-p}
\eea
\noindent The time-dependent part $P_1(t)$ is solely determined in terms of the non-unitary matrix $U$ and has the expression:
\bea
P_1(t) =  b_0 \left [ 1 + \frac{2 \Gamma D}{\epsilon_0^2} 
\sin^2(\epsilon)
 + \left ( \frac{b_3 \Gamma}{b_0 \epsilon_0} \right ) \sin(2\epsilon) \right ]
\label{exp-p1}
\eea
\noindent The expression of $P_1(t)$ for constant $k$ and $\gamma$ for which $\mu(t)=t,
\epsilon= \epsilon_0 t $ has been obtained earlier\cite{pkg}. The
effect of allowing $k$ and $\gamma$ to have identical time dependence specified by the common
scale factor $\mu_0(t)$ is solely contained in the argument of the sine function. The power-oscillation
vanishes for $b_3=0, D=0$ which is also the limit for time-independent $W^{\dagger} F_{\pm} W=\frac{b_0}{2}$. 

\section{Systems without external potential $V(x)$ }

We have described the general method for obtaining analytic solutions of Eq. (\ref{nlse1}) 
in section III. In this section, we present some specific examples for $V(x) = 0$ by considering
time-independent and time-dependent LC and BLG terms separately. The case of non-vanishing $V(x)$
will be considered in the next section.  

\subsection{Time-independent LC and BLG terms}

We consider time-independent LC and BLG terms for which $\mu_0$,
$k=\mu_0 \beta$ and $\gamma=\mu_0 \Gamma$ are constants. The
expression of the non-unitary matrix $U(t)$ may be obtained by adjusting
Eqs. (\ref{mu1}) and Eq.(\ref{U}) as,
\bea
U(t) = \sigma_{0} \cos(\epsilon_0 t) - \frac{i A_{0}}{\epsilon_0} \sin(\epsilon_0 t) 
\eea 
\noindent where we have chosen $\mu_0=1$ without loss of any generality. The
time-dependent function $\epsilon(t)$ appearing in the expression of $P_1$
in Eq. (\ref{exp-p1}) takes the form $\epsilon(t) = \epsilon_0 t$. The space-time
modulations of the nonlinear strength are discussed by considering (i) constant,
(ii) purely time-dependent, (ii) purely space dependent and (iv) space-dependent
$g_{ij}$ separately.

\subsubsection{ Constant $g_{ij}$}

The functions $g_{ij}$ in Eq. (\ref{geqn-3}) are space-time independent provided $f_1, f_2, G$
and $W^{\dagger} F_{\pm} W$ are chosen to be constants. It may be recalled that $W^{\dagger}
F_{\pm} W = \frac{b_0}{2}$ is constant  for $W_{1} = W_{2}$ and $D=0$, i.e. $\Gamma = -
{\vert \beta \vert} \sin \left (\theta_{2}- \theta_{1} -\theta_{3} \right )$.
The expressions of $g_{ij}$ for these choices are  obtained as,
\bea
g_{11} & = & \frac{1}{b_0}(f_{1}+G) \ ; \
g_{21}= \frac{1}{b_0}(f_{1}-G) \nonumber \\
g_{22} & = & \frac{1}{b_0}(f_{2}+G) \ ; \
g_{12}   =  \frac{1}{b_0}(f_{2}-G)
\label{G1}  
\eea
\noindent where the constants $f_1, f_2, G$ may be chosen independently for describing
different physical situations. For example, the choice $f_1=f_2$ leads to $g_{11}=g_{22}\equiv g,
g_{12}=g_{21} \equiv \tilde{g}$. The system admits a Lagrangian-Hamiltonian formulation for
$g_{12}=g_{21}$ and the relevant expressions are given in Eq. (\ref{lang-hami}).
In the terminology of optics, $g$ and $\tilde{g}$ may be identified  as self-phase
modulation and cross-phase modulation, respectively. The self-phase and cross-phase modulation
terms may be made vanishing by choosing $f_1=f_2=G$ and $f_1=f_2=-G$, respectively.
The nonlinear interaction becomes $SU(2)$ invariant for the choice $f_1=f_2, G=0$ for
which $g=\tilde{g}$. This particular system has been studied earlier in the context of optics
and to the best of our knowledge, no exact solution has been found for generic values of
$g_{ij}$. We present below exact analytical solutions for arbitrary $g_{ij}$
for the first time, which automatically includes the specific values of $g_{ij}$ discussed above. 

With the choice of $f=\frac{1}{2} (f_1+f_2) \equiv 2 \sigma$ and $V(x)=0$,
Eq.(\ref{R3}) reduces to, 
\bea
R_{xx} + E  R - \frac{C^2}{R^3} - 2\sigma R^{3} = 0,
\label{R6}
\eea
\noindent which is exactly solvable. Hence, no further transformation as in Eq.(\ref{zeta}) is required. The exact
solutions of Eq. (\ref{R6}) is discussed in Appendix-I. We denote the solutions of Eq. (\ref{R6}) for $C=0, E <0,
\sigma <0$ as $R_0(x)$ with its analytical expression given by,
\bea
R_0(x) = \sqrt{\frac{\vert E \vert}{\vert \sigma \vert}} \sech \left(\sqrt{\vert E \vert} x\right )
\label{R7.0}
\eea
\noindent The solution describes a soliton and has been obtained earlier\cite{Driben2}. We present new solutions by
taking $C \neq 0$ and $\sigma \geq 0$. For $\sigma =0$, i.e. $f_1=-f_2$, Eq. (\ref{R6}) is the famous Ermakov-Pinney
equation
and admits periodic solution $R_P(x)$ for $E > 0$,
\bea
R_P(x) & = & \sqrt{\sin^2(\sqrt{E} x )+\frac{C^2}{E}\cos^2(\sqrt{E} x)}, \ E  > 0.
\eea
\noindent while diverges for $E < 0$. The solutions of Eq. (\ref{R6}) for $\sigma > 0$ are expressed in terms
of Weierstrass Elliptic functions and choosing $\sigma = 1$ for simplicity, one of the solutions reads,
\bea
R(x) =\sqrt{ Q_1 + (Q_2 - Q_1) sn^{2}(\lambda x, r)},
\label{R7}
\eea 
\noindent where $\lambda = \sqrt{Q_3 - Q_1}$, $r^2 = \frac{Q_2-Q_1}{Q_3-Q_1}$ and $Q_1,Q_2,Q_3$ are constants depending
on the arbitrary real constants $E$, $C$, $\sigma$. Thus, the constants $Q_i$'s may be chosen conveniently by fixing
the values of $E, C, \sigma$. The solution is chosen to be bounded as $0 < Q_1 \leq R^{2} \leq Q_2 <
Q_3 $. In the limit $Q_3 \rightarrow Q_2$, the solution reduces to
\bea
R(x) = \sqrt{ Q_1 + (Q_2 - Q_1) \tanh^{2}(\lambda x)}.
\label{R7.1}
\eea
\noindent Following the discussions in Appendix-I, other solutions may be written down easily.
The power $P$ of the system for constant $g_{ij}$ is time independent, i.e. $P=b_0 R^2$, where
$R$ is given by Eq. (\ref{R7}) or in the limit $Q_3 \rightarrow Q_2$ by Eq. (\ref{R7.1}). The power-oscillation
in time has been observed for models described in Ref. \cite{pkg} where the nonlinear potential
is of the form $(\Psi^{\dagger} M \Psi)^2$ where the scalar gauge potential $A$ is pseudo-hermitian
with respect to the constant matrix $M$, i.e.  $A^{\dagger} = M A M^{-1}$. The nonlinear potential
defined by $g_{ij}$'s in Eq. (\ref{G1}) can not be cast in the form  $(\Psi^{\dagger} M \Psi)^2$ for some
$M$ and the absence of power oscillation for the present case is not in  contradiction with the results of Ref.
\cite{pkg} .
 
\subsubsection{Purely time-dependent $g_{ij}$} 

We consider Eq.(\ref{nlse1}) with purely time-dependent nonlinear strengths
by choosing $g_{ij}$'s to be dependent on time alone. It may be recalled that
time dependent non-linear strength arises in optics for transverse beam propagation
in layered optical media\cite{Berge,Towers}. Further, NLSE with periodic time variation
of nonlinear strength in an external magnetic field has been used to produce matter wave
breathers in quasi one-dimensional Bose-Einstein condensate\cite{Kevrekidis2}.
As evident from Eq. (\ref{geqn-3}), purely time dependent $g_{ij}$'s may be obtained by taking
$f_1, f_2$ as constants and allowing $G$ to be a function of time only. For simplicity,
we consider $\theta_2-\theta_1-\theta_3 = n\pi$ and $W_1 = W_2$, for which the expressions of $g_{ij}$ are,
\bea
g_{11} & = & \frac{f_{1}+G(t)}{b_{0}(1+ 2 \Gamma q_{+})}, \ \
g_{21}= \frac{f_{1}-G(t)}{b_{0}(1+2 \Gamma q_{+})}, \nonumber \\
g_{22} & = & \frac{f_{2}+G(t)}{b_{0}(1+2 \Gamma q_{-})}, \ \
g_{12}   =  \frac{f_{2}-G(t)}{b_{0}(1+ 2 \Gamma q_{-})},
\label{G2} 
\eea
\noindent where $q_{\pm}$ are given as,
\bea
q_{\pm} = \frac{\Gamma}{\epsilon^{2}_{0}} \sin^{2}(\epsilon_0 t) \pm \frac{\sin(2 \epsilon_0 t)}{2 \epsilon_0}.
\eea
\noindent The functions $q_{\pm}$ introduce periodic time-modulation of the nonlinear strengths.
Eq.(\ref{R3}) reduces to Eq.(\ref{R6}) for this case also, since $f_1, f_2$ are constants and $V(x)=0$.
Consequently, $R(x)$ have the expressions given by Eqs. (\ref{R7.0}), (\ref{R7}) and (\ref{R7.1}) for the
specified choices of the constants. Unlike the case of constant $g_{ij}$,
the power is time-dependent,
\bea
P  =  b_0 \left [ 1 + \frac{2 \Gamma^{2}}{\epsilon_0^2} \sin^2(\epsilon_0  t) \right ]  R^{2}(x) 
\eea
\noindent where any one of the expressions of $R$ given in Eqs. (\ref{R7.0}), (\ref{R7}) and (\ref{R7.1}) may be
used within their ranges of validity.

\subsubsection{Purely space-dependent $g_{ij}$}

It is evident from Eq. (\ref{geqn-3}) that purely space-dependent $g_{ij}$'s are obtained 
for constant $W^{\dagger} F_{\pm} W$ and choosing purely space dependent $G(x,t)\equiv G(x)$.
As discussed in Sec. III,  $W^{\dagger} F_{\pm} W = \frac{b_0}{2} $ is constant in the limit
$W_{1} = W_{2}$ and $D=0$. The space modulation is determined by the space-dependent
functions $f_1(x), f_2(x)$ and $G(x)$. Unlike the previous cases, solution of Eq. (\ref{R3})
for arbitrary $f(x)$ necessitates the transformation given by Eq. (\ref{zeta}) resulting in
Eqs. (\ref{u1},\ref{u1.5}).  The solutions of Eq.(\ref{u1.5}) for $V(x)=0$ are,
\bea
\rho^2(x) & = & a e^{2\sqrt{\vert E \vert} x} + \frac{m}{4 \vert E \vert a} e^{-2\sqrt{\vert
E \vert} x}, \ \ E<0  \nonumber\\
\rho^2(x) & = & a \sin^2(\sqrt{E} x) + \frac{m}{a E} \cos^2(\sqrt{E} x), \ \  E>0 
\label{rho-exp}
\eea
\noindent where the constant $a > 0$. The solutions for $E<0$  diverges for large $x$ and
will not be considered further in this article. We consider $E > 0$ and make the following
choices for a simplified expression of $\rho$, $E=  \frac{\omega^2}{4}, a =(1-\alpha),
m = \frac{\omega^2(1-\alpha^2)}{4}$, where $\omega$ and $\alpha$ are real constants.
The expressions of $\rho^2(x)$ and hence, $f(x)$ have the following simplified expressions,
\bea
\rho^{2}(x) = 1 + \alpha \cos(\omega x), \ \
f(x) = \frac{2 \sigma}{ [ 1 + \alpha \cos(\omega x)]^{3}}.
\label{rho-f}
\eea
\noindent The functions $f_1(x)$ and $f_2(x)$ may be chosen independently subjected to
the constraint $f(x)=\frac{1}{2}(f_1(x) + f_{2}(x))$. We choose $f_1(x)=f_2(x)= 2
\sigma [ 1 + \alpha \cos(\omega x)]^{-3}$ for which space-dependent $g_{ij}$ have the
expressions,
\bea
&& g_{11} = g_{22}=\frac{G(x)}{b_0} + \frac{2 \sigma }{b_0 [ 1 + \alpha \cos(\omega x)]^{3}},
\nonumber \\
&& g_{12}=g_{21} = -\frac{G(x)}{b_0} + \frac{2 \sigma }{b_0 [ 1 + \alpha \cos(\omega x)]^{3}}
\eea
\noindent For $G=0$, all $g_{ij}$'s become identical. 
The expression of $\zeta$ is obtained by using the second equation of Eq. (\ref{zeta})
and the first equation of (\ref{rho-f}),
\bea
\zeta = \frac{2}{\omega\sqrt{1-\alpha^2}} \ \arctan \left [ \sqrt{\frac{1-\alpha}{1+\alpha}} \
\tan(\frac{\omega x}{2}) \right ],
\label{zeta-exp}
\eea
\noindent where ${\vert \alpha \vert} < 1$. It remains to find $u(\zeta)$ in order to completely
specify the solution. Note that the equations (\ref{R6}) and  (\ref{u1})  satisfied by $R(x)$
and $u(\zeta)$, respectively, are identical with the identification of
$ E \leftrightarrow m, x \leftrightarrow \zeta, R(x) \leftrightarrow u(\zeta)$. The solutions
of Eq. (\ref{R6}) are given in Eqs. (\ref{R7.0}) and (\ref{R7}) which may be used to
write down the solutions of $u(\zeta)$ with the identification stated above.  
The power $P = b_0 \rho^{2}(x) u(\zeta)^{2}$ is independent of time.

\subsubsection{Space-time dependent $g_{ij}$}

The results for space-time dependent $g_{ij}$'s have been discussed in Sec. III in detail except 
for the solution of Eq. (\ref{R3}) for $R(x)$ or equivalently of Eq. (\ref{u1},\ref{u1.5}) for $u(\zeta)$ 
and $\rho(x)$. It may be noted that Eq. (\ref{u1}) is the same for purely space dependent $g_{ij}$ as 
well as space-time dependent $g_{ij}$. Thus, the expressions of
$\{ \rho(x), f(x) \}$ and $\zeta(x)$ are given by Eqs. (\ref{rho-f}) and (\ref{zeta-exp}), respectively, 
within the specified ranges of validity. The solutions for $u(\zeta)$ are given by Eqs. (\ref{R7.0}) 
and (\ref{R7}) with the replacement of $ E \rightarrow m, x \rightarrow \zeta, \{ R_0(x), R_P(x), R(x) \} 
\rightarrow \{ u_0(\zeta), u_P(\zeta, u(x) \}$. The exact solution $\Psi(x,t)$ for space-time dependent 
$g_{ij}$ is different from purely space-dependent $g_{ij}$ due to the time-dependence. In particular, 
the power is time-independent for purely space-dependent $g_{ij}$, while it is time-dependent for 
space-time dependent $g_{ij}$ and given by Eqs. (\ref{exp-p}) and (\ref{exp-p1})
with $\epsilon=\epsilon_0 t$ and $\mu=1$. 

\subsection{Time-dependent $k, \gamma$}

The equations determining $R(x)$ and $\theta(x)$ are the same as in the case of constant $k, \gamma$.
However, the non-unitary matrix $U(t)$, the strengths $g_{ij}$, and hence, expression of power
will change depending on specific forms of $k(t)$ and $\gamma(t)$. We have already discussed
solutions for $R(x)$ and $\theta(x)$ depending on constant, purely time-dependent, purely
space-dependent and space-time dependent $g_{ij}$ which are valid for time-dependent $k(t),
\gamma(t)$ independent of specific time-dependence. In order to avoid repetition of the same
results, we present results related to $U(t), g_{ij}(t), P(t)$ for specific time-dependence of
$k(t)$ and $\gamma(t)$. We choose periodic modulation of the LC and BLG terms by choosing
$\mu_0(t)=\cos(\omega_0 t), \omega_0 \in \mathbb{R}$ so that,
\bea
&& \mu(t)=\frac{1}{\omega_0} \sin(\omega_0 t), \
\epsilon=\frac{\epsilon_0}{\omega_0} \sin(\omega_0 t), \ 
T_{\pm}(t)=\frac{\omega_0 p_{\pm}}{\sin(\omega_0 t)},\nonumber \\ 
&& p_{\pm} \equiv \frac{\Gamma}{\epsilon^{2}_{0}} \sin^{2} \left (
\frac{\epsilon_0}{\omega_0} \sin(\omega_0 t) \right ) \pm
\frac{1}{2\epsilon_0}\sin \left (\frac{2\epsilon_0}{\omega_0}
\sin(\omega_0 t) \right ).\nonumber
\eea
\noindent The limit $\omega_0 \rightarrow 0$ corresponds to constant
LC and BLG terms. The expressions for $U, W^{\dagger} F_{\pm} W, g_{ij},
P$ may be obtained from Eqs.  (\ref{U}), (\ref{wpmw}), (\ref{geqn-3}) and (\ref{exp-p1}),
respectively by using the above expressions for $\mu(t), \epsilon(t), T_{\pm}(t)$.

The expressions of $W^{\dagger} F_{\pm} W$,
\bea
W^{\dagger} F_{\pm} W & = & \frac{b_0}{2} \left ( 1 + 2 D p_{\pm} \right )
+ \frac{b_3}{2} \left [ \frac{\Gamma}{\epsilon_0} \sin\left ( \frac{2 \epsilon_0}{\omega_0}
\sin(\omega_0t) \right )
 \pm \cos \left ( \frac{2 \epsilon_0}{\omega_0} \sin(\omega_0t) \right )\right ],
\eea
\noindent when substituted in Eq. (\ref{geqn-3}) gives the space-time dependent $g_{ij}$ for
periodic time-modulation of the LC and BLG terms. Different types of space-time modulations of
the nonlinear strengths $g_{ij}$ may be considered as follows:
\begin{itemize}
\item Constant $g_{ij}$: $D=0, b_3=0, (G, f_1, f_2)=\textrm{constant}$
\item Purely time-dependent $g_{ij}$: Either $D$ or $b_3$ or both $D$ and $b_3$ are non-vanishing;
$f_1$ and $f_2$ are constants and $G(x,t) \equiv G(t)$
\item Purely space-dependent $g_{ij}$: $D=0=b_3$, $G(x,t)\equiv G(x)$
\end{itemize}
The condition $D=0$ may be imposed by using Eq. (\ref{D-eqn}) which gives a specific relation
among the parameters, while $b_3=0$ for the choice $W_1=W_2$. The functions $f_1(x), f_2(x),
G(x,t)$ may be chosen as per the requirement, since they are arbitrary. The non-unitary operator
$U$ and $P$ have the expressions:
\bea
U & = & \sigma_0 \cos \left (\frac{\epsilon_0}{\omega_0} \sin(\omega_0 t) \right )
- \frac{i A_0}{\epsilon_0} \sin \left ( \frac{\epsilon_0}{\omega_0} \sin(\omega_0 t) \right )\nonumber \\
P (x,t) & = & b_0 \left [ 1 + \frac{2 \Gamma D}{\epsilon_0^2} \sin^2 \left
( \frac{\epsilon_0}{\omega_0} \sin(\omega_0 t) \right ) 
 + \frac{b_3 \Gamma}{b_0 \epsilon_0}
\sin \left ( \frac{2 \epsilon_0}{\omega_0} \sin(\omega_0 t) \right ) \right ] R^2(x),
\eea
\noindent appropriate expressions for $R(x)$ are to be substituted for a given $g_{ij}$.

\begin{figure}[H]
	\begin{subfigure}{0.32\textwidth}
		\centering
		\includegraphics[width=0.99\linewidth]{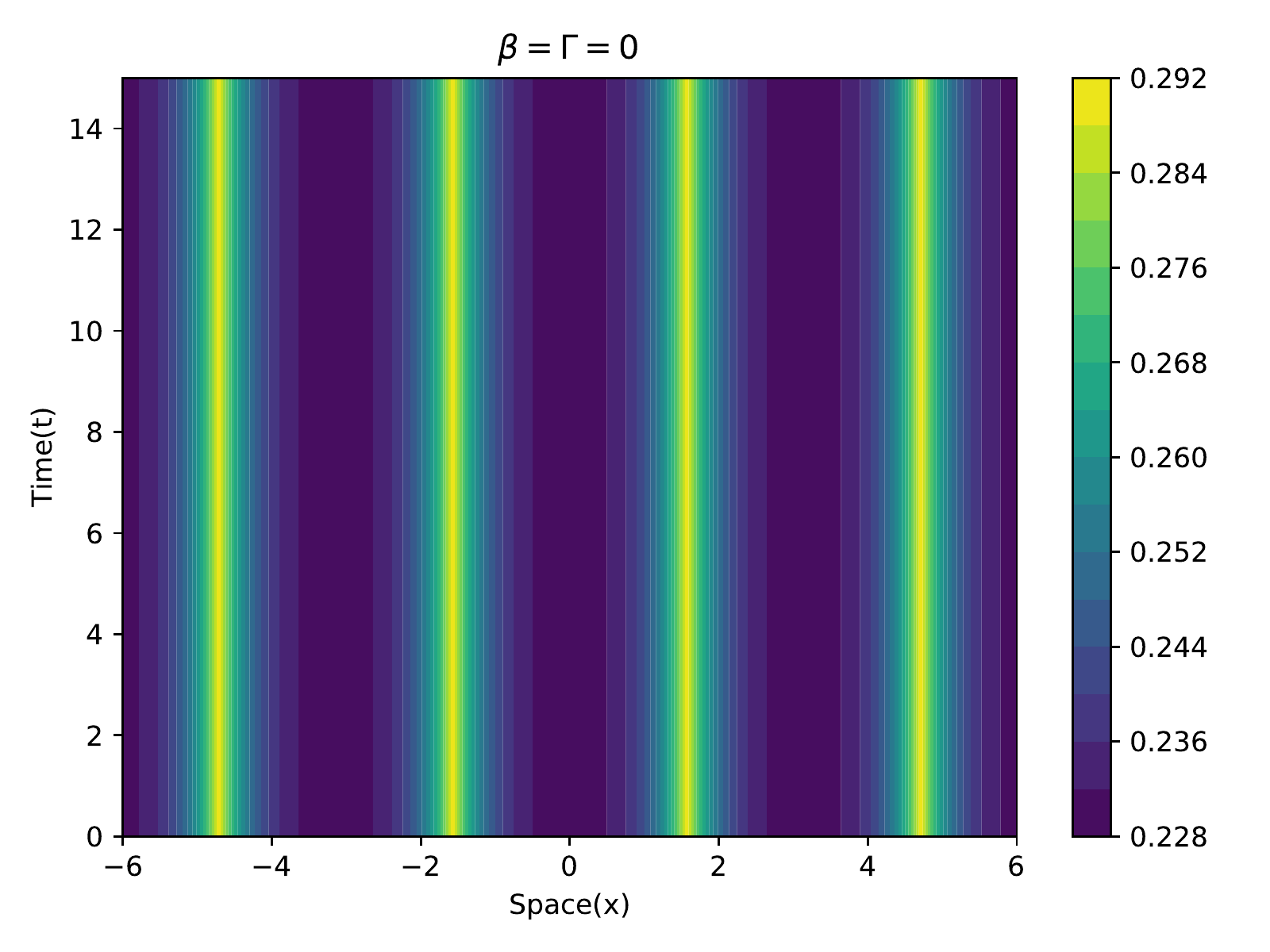}\quad
		\caption{}
		\label{fig:1}
	\end{subfigure}
	\begin{subfigure}{0.32\textwidth}
                \centering
                \includegraphics[width=0.99\linewidth]{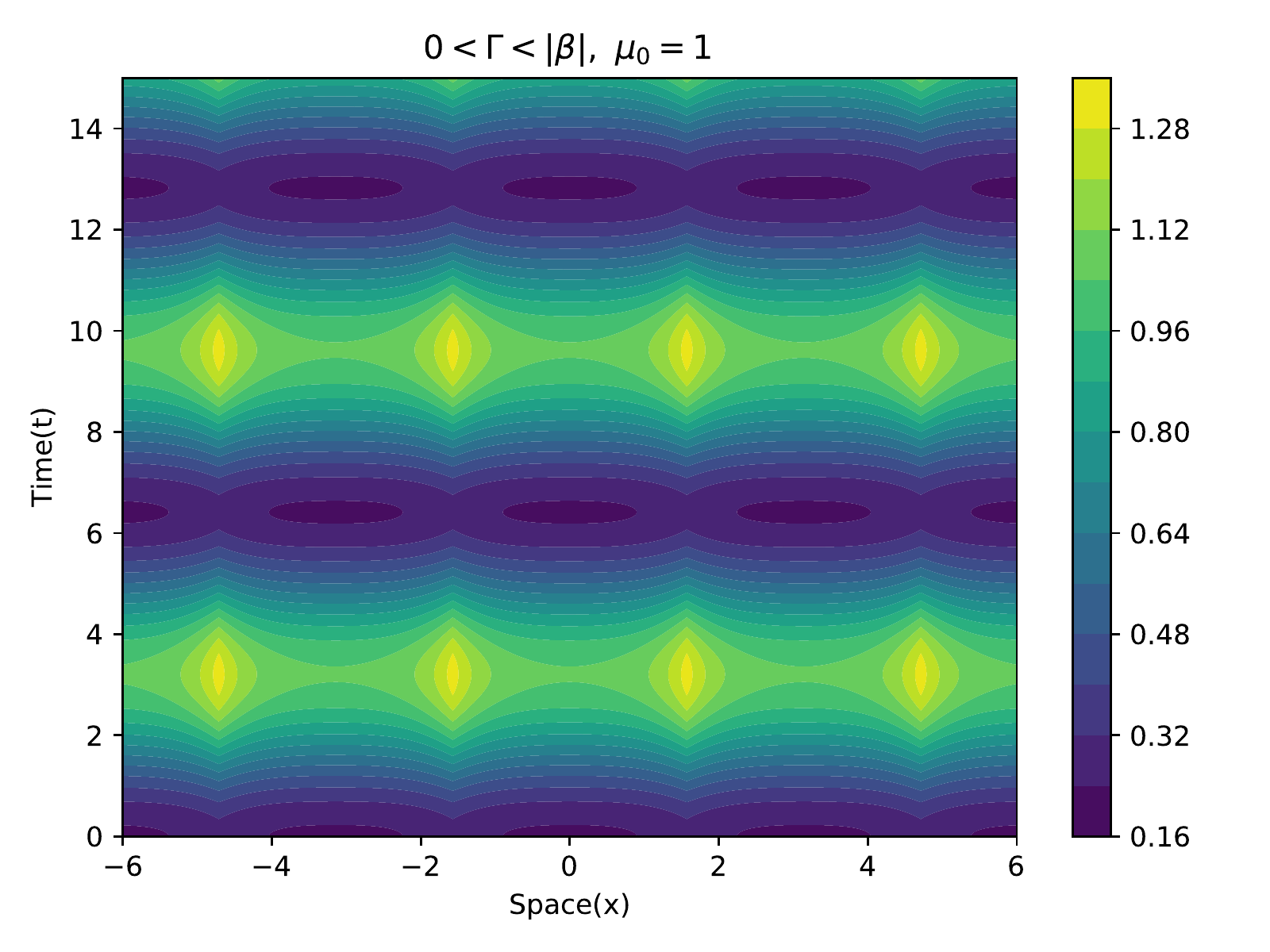}\quad
                \caption{}
                \label{fig:2}
        \end{subfigure}
	\begin{subfigure}{0.32\textwidth}
                \centering
                \includegraphics[width=0.99\linewidth]{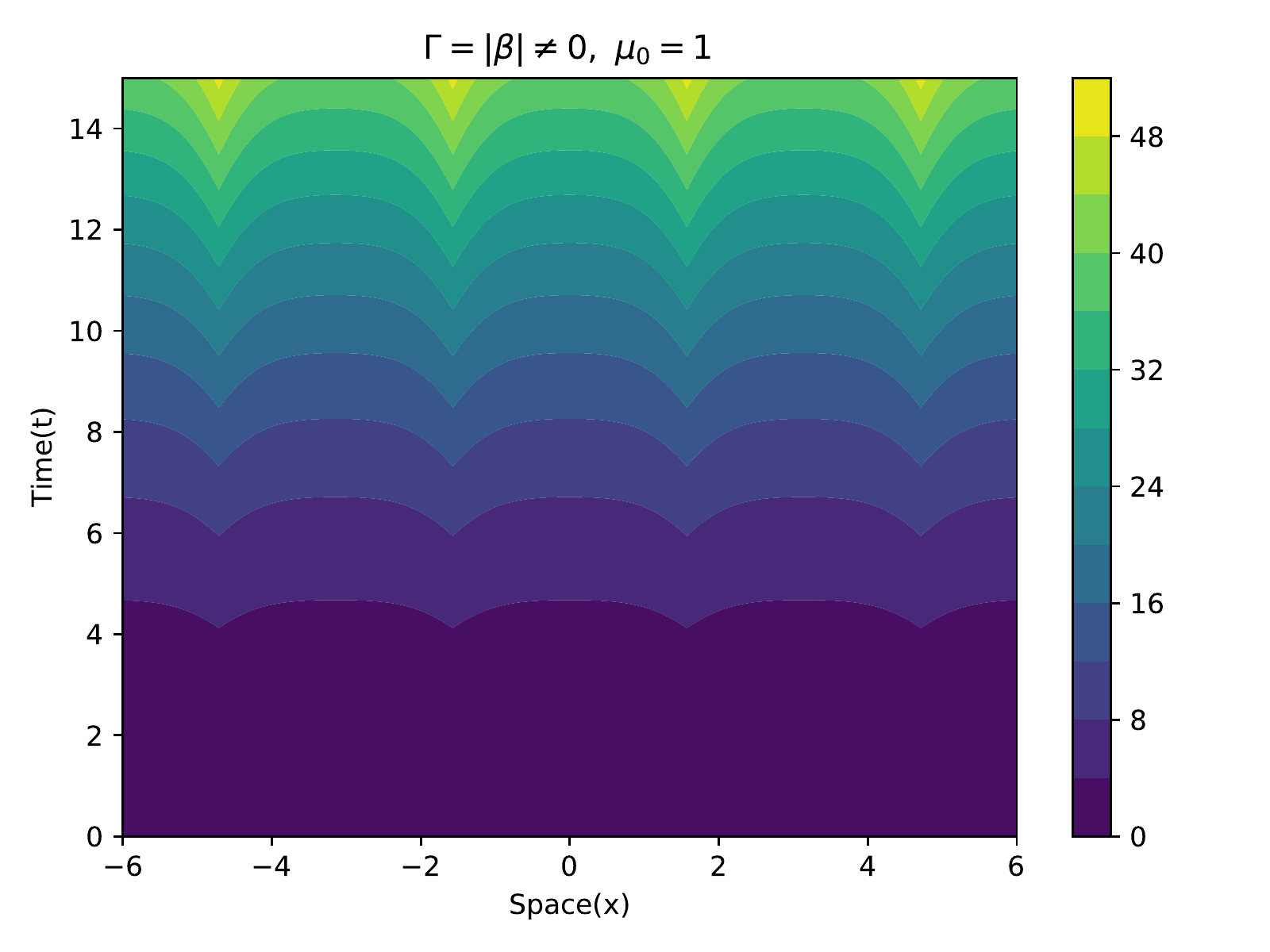}\quad
                \caption{}
                \label{fig:3}
        \end{subfigure}
	\medskip
	\begin{subfigure}{0.32\textwidth}
                \centering
                \includegraphics[width=0.99\linewidth]{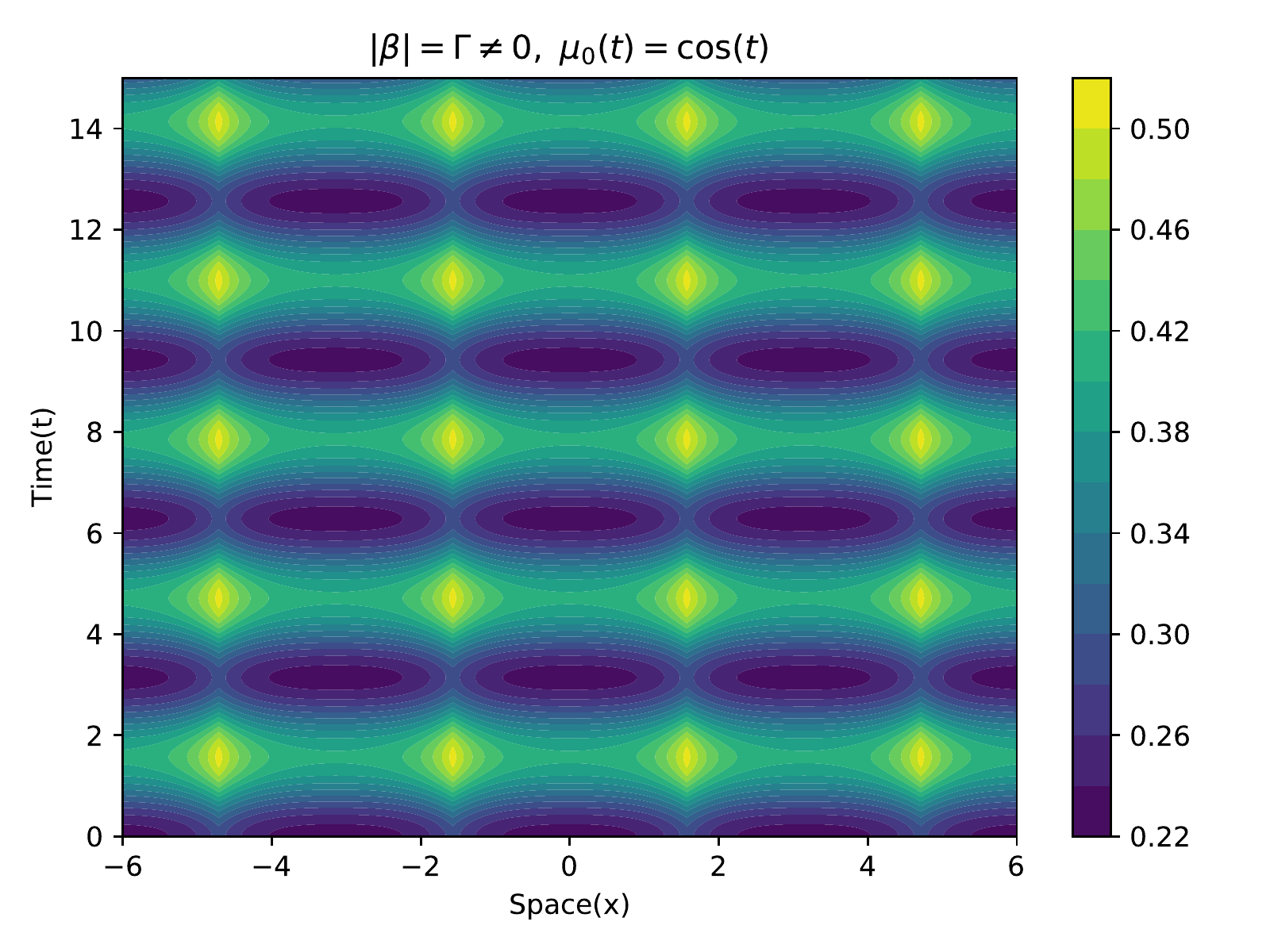}\quad
                \caption{}
                \label{fig:4}
        \end{subfigure}
	\begin{subfigure}{0.32\textwidth}
                \centering
                \includegraphics[width=0.99\linewidth]{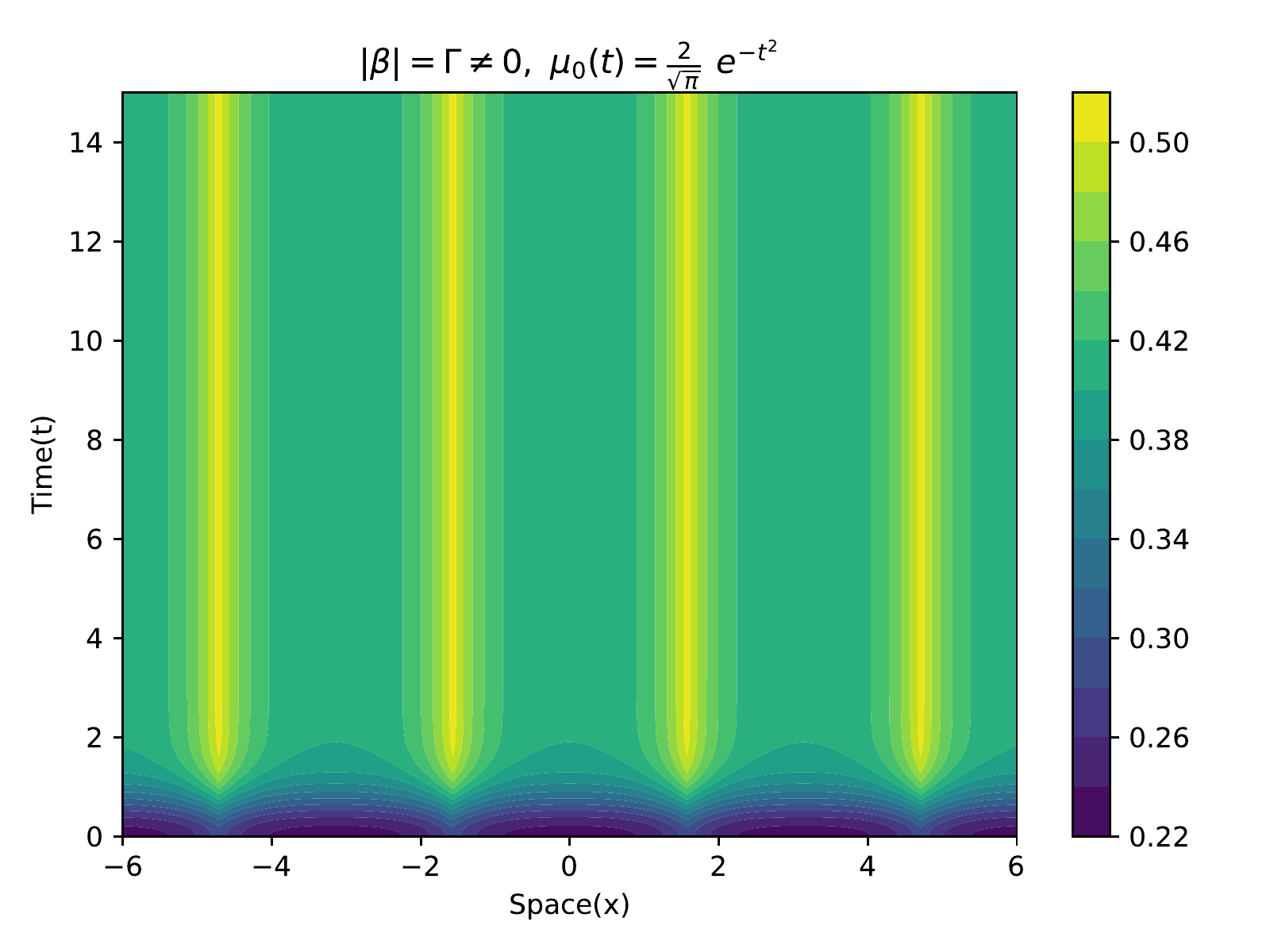}\quad
                \caption{}
                \label{fig:5}
        \end{subfigure}
	\begin{subfigure}{0.32\textwidth}
                \centering
                \includegraphics[width=0.99\linewidth]{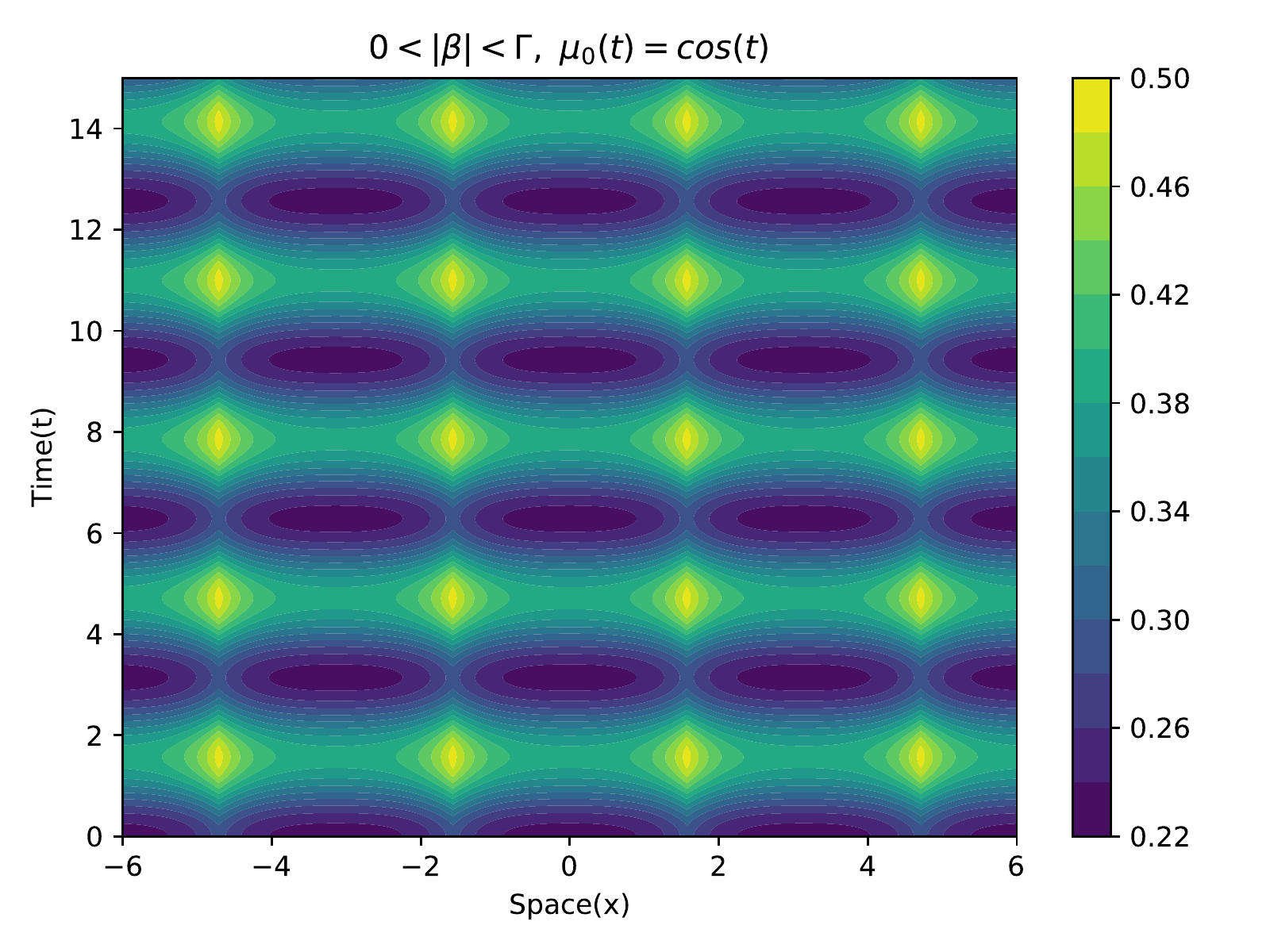}\quad
                \caption{}
                \label{fig:6}
        \end{subfigure}
	\caption{ (Color online) Plot of Power i.e $\Psi^{\dagger}\Psi$ for $V(x) = 0$ and the expression 
	of $\Psi$ is shown in Eq. (\ref{sol1}). Parameters: 
	$W_1=W_2=1,\theta_1 = \frac{\pi}{6}, \theta_2 = \frac{2\pi}{3}, \theta_3 = \frac{\pi}{3},
	Q_1 = 0.1, Q_2 = 0.29, Q_3 = 0.3, \sigma = 1, \alpha = 0.15, \omega =2$. In Fig(a):
	$\Gamma = |\beta| =0$, Fig(b): $\Gamma = 0.5, |\beta| =0.7, \mu_0 = 1$, Fig(c): $\Gamma = |\beta|=0.5,
	\mu_0 =1$, Fig(d): $\Gamma=|\beta|=0.5,\mu_0(t) = cos(t)$, Fig(e): $\Gamma=|\beta|=0.5, \mu_0(t) 
	= \frac{2}{\sqrt{\pi}} e^{-t^2}$, Fig(f): $\Gamma=0.5,|\beta|=0.3,\mu_0(t) = cos(t)$}
	\label{img1}
\end{figure}

The general explicit solution of Eq. (\ref{nlse1}) for $V(x) = 0$ and space-time 
dependent $g_{ij}$ is 
\bea
\psi_1(x,t) &  =     & \left[ W_1 e^{i\theta_1} \cos(\epsilon) - \left\{ i |\beta| W_2 
e^{i\left(\theta_2 -\theta_3\right)} - \Gamma W_1 e^{i\theta_1} \right\} 
\frac{\sin(\epsilon)}{\epsilon_0} \right] \sqrt{1+\alpha \cos(\omega x)} \nonumber \\
            & \times & \sqrt{Q_1+ (Q_2 - Q_1) sn^{2}(\lambda \zeta,r)} \ 
	    e^{i\left(\theta(x)-Et\right)}\nonumber \\
\psi_2(x,t) &  =     & \left[ W_2 e^{i\theta_2} \cos(\epsilon) - \left\{ i |\beta| W_1 
e^{i\left(\theta_1 +\theta_3\right)} + \Gamma W_2 e^{i\theta_2} \right\} 
\frac{\sin(\epsilon)}{\epsilon_0} \right] \sqrt{1+\alpha \cos(\omega x)} \nonumber \\
            & \times & \sqrt{Q_1+ (Q_2 - Q_1) sn^{2}(\lambda \zeta,r)} \ 
	    e^{i\left(\theta(x)-Et\right)} 
\label{sol1}
\eea
The expression of $\zeta$ is given in Eq. (\ref{zeta-exp}) and $\theta(x)$ can be obtained 
from Eq. (\ref{theta}). The power $P = \Psi^{\dagger}\Psi$ is plotted in FIG. \ref{img1} for 
various time-modulations. The case of vanishing BLG and LC is considered Fig. 1(a) and
no variation of the power with time is seen. The power-oscillation is seen in Fig. 1(b) for
constatnt $\mu_0$ and non-vanishing $\beta, \Gamma$ satisfying ${\vert \beta \vert} > \Gamma$.
The power-oscillation in time is a manifestation of the balanced loss-gain in the system. The
solution $\Psi$ becomes unstable for non-vanishing $\beta, \Gamma$ satisfying $|\beta| \leq \Gamma$ and 
constant $\mu_0$. The respective plot is given in Fig 1.(c). We can manage the instability by choosing 
$\mu_0(t)$ properly. In Fig. 1(d), we have considered $\vert \beta \vert = \Gamma$ and a periodic 
modulation function $\mu_0(t)=\cos(t)$ for which $P(x,t)$ shows periodic behaviour. We have plotted 
power for $\vert \beta \vert = \Gamma$ and $\mu_0(t) = \frac{2}{\sqrt{\pi}} e^{-t^2}$ in Fig. 1(e) 
which is also bounded in time. The instability in the region $\vert \beta \vert < \Gamma$ can again
be controlled by periodic modulation. In particular, the plot in Fig. 1(f) shows power oscillation
for $\mu_0(t)=\cos t $ in the region $\vert \beta \vert < \Gamma$. The figures are 
shown upto time $t \approx 15$, but the stated features have been checked upto time $t \approx 200$.

\section{Systems with Real $V(x)$ }

In this section, results for non-vanishing real potential $V(x)$ will be presented for purely
space-dependent and space-time dependent $g_{ij}$. The method outlined in Sec. III is not
suitable for constant or purely time-dependent $g_{ij}$ for which $f_1, f_2$ are necessarily
constants. Consequently, it follows from Eq. (\ref{u1.5}) that $f(x), \rho$ and $s(x)$ are
also constants for real $V(x)$. Further, a direct solution of Eq. (\ref{R3}) for constant $f(x)$ and
spatially varying $V(x)$ is also unknown. Within this background, we discuss a few examples
of real $V(x)$ for space-time dependent $g_{ij}$ and the limit of purely space-dependent $g_{ij}$
may be considered by choosing $D=0=b_3$ and $G(x,t)\equiv G(x)$. The sole effect of adding a
non-vanishing real potential $V(x)$ to a system with $V(x)=0$ is modified expressions of
$\rho(x)$ and hence, of $\zeta(x)$ and $f(x)$. The equation determining
$u(\zeta)$ remains unaffected by the choice of $V(x)$, although $u(\zeta(x))$ is dependent
on the choice of $V(x)$ via $\zeta(x)$. The knowledge of $f(x)$ specifies the space-dependent
part of $g_{ij}$\footnote{$G(x,t)$ appearing in the expressions of $g_{ij}$ does not
affect any results and may be chosen to be zero}, while space-dependent part of the power
$P$ is modified due to the expression
of $R(x)$ which depends on $\rho(x)$ and $u(\zeta(x))$. We present the expressions of $\rho(x),
f(x), \zeta(x), u(\zeta)$ for a given $V(x)$ and do not repeat the results and discussions which
have been presented in earlier sections. The space-modulated nonlinear strengths are determined
in terms of $f_1(x)$ and $f_2(x)$. One may choose $f_1=f_2=f$ for simplicity or choose
$f_1$ and $f_2$ such that $f(x)=\frac{1}{2} (f_1(x)+f_2(x))$ is satisfied.  
There is a freedom in tailoring the nonlinear strengths $g_{ij}$ for a given $f(x)$ which
may be exploited to construct physically interesting systems.

\subsection{Reflection-less Potential}

We choose the reflection-less potential,
\bea
V(x) = N^2 - N(N-1) sech^{2}(x), \ N \in \mathbb{Z}^> .
\eea
\noindent The solutions for $\rho(x)$ and $f(x)$ for $E = m = 0$ may be obtained
from Eq.(\ref{u1.5}) as,
\bea
\rho(x) = \cosh^{N}(x), \ \ f(x) = 2 \sigma sech^{6N}(x).
\eea
\noindent The potential $V(x)$ has been considered earlier in  the context of
single component local\cite{Beitia1} as well as non-local\cite{pkg1} NLSE.
The nonlinear strengths $g_{ij}$'s may be tailored for the above $f(x)$ satisfying
$\frac{1}{2} (f_1(x)+f_2(x))=f(x)$. 
Following the discussions in Appendix-I, the sum of the roots $Q_1,Q_2,Q_3$ is zero
for $m=0$, implying that all three roots can not either be positive or negative. We choose $Q_1 <0,
Q_2 >0, Q_3 >0$. The solution of Eq.(\ref{u1}) for $\sigma=-1$ is obtained as,
\bea
u(\zeta) = \sqrt{Q_3 - (Q_3 - Q_2) sn^2(\lambda \zeta,r)}
\label{uzeta}
\eea
\noindent where $\lambda = \sqrt{Q_3 - Q_1}$ and $ r^2 = \frac{Q_3 - Q_2}{Q_3 - Q_1}$.
The expression of $\zeta$ is obtained from Eq. (\ref{zeta}),
\bea
\zeta = \int \sech^{2N}(x) dx,
\eea
\noindent which can be evaluated for fixed $N$ by  using the formula,
$\int sech^{n}(x) = \frac{sech^{n-2}(x) tanh(x)}{n-1} + \frac{n-2}{n-1} \int sech^{n-2}(x) $ 
repeatedly.

\subsection{Quadratic Potential}

We consider the quadratic potential $V(x) = E^2 x^2$ and $E > 0$. For $m = 0$, the solution 
of $\rho(x)$ and $f(x)$ is obtained from Eq.(\ref{u1}) as,
\bea
\rho(x) =  e^{-\frac{1}{2} E x^2}, \ \
f(x) =  2 \ \sigma \ e^{3 E x^2}.
\eea
\noindent An unpleasant feature is that the nonlinear strengths $g_{ij}$ grow in space
for this specific $f(x)$. The freedom in choosing $f_1(x), f_2(x)$ satisfying
the constraint $f(x)=\frac{1}{2}(f_1(x)+f_2(x))$ and arbitrary function $G(x,t)$ is of no
help to construct localized $g_{ij}$'s due to their specific dependence on $f_1, f_2, G$.
The solutions for $u(\zeta)$ is same as (\ref{uzeta}) with 
the expression of $\zeta$ obtained from Eq.(\ref{zeta}) as,
\bea
\zeta = \frac{1}{2} \sqrt{\frac{\pi}{E}} \ erfi(\sqrt{E} x),
\eea  
\noindent where $erfi$ is imaginary error function. The function $\Psi(x,t)$ is finite in all 
regions of space, since $R(x) \rightarrow 0$ as $|x| \rightarrow \infty $. On other hand, 
$\Psi(x,t)$ diverges for $E<0$.

The general explicit solution of Eq. (\ref{nlse1}) for Quadratic Potential and space-time
dependent $g_{ij}$ is
\bea
\psi_1(x,t) &   =    & \left[ W_1 e^{i\theta_1} \cos(\epsilon) - \left\{ i |\beta| W_2 
e^{i\left(\theta_2-\theta_3\right)} - \Gamma W_1 e^{i\theta_1} \right\} 
\frac{\sin(\epsilon)}{\epsilon_0} \right] \ e^{-\frac{1}{2} E x^2} \nonumber \\
            & \times & \sqrt{Q_3 - (Q_3 - Q_2) sn^{2}(\lambda \zeta,r)} \ 
	   e^{i\left(\theta(x)-Et\right)} \nonumber \\
\psi_2(x,t) &  =     & \left[ W_2 e^{i\theta_2} \cos(\epsilon) - \left\{ i |\beta| W_1 
e^{i\left(\theta_1+\theta_3\right)} + \Gamma W_2 e^{i\theta_2} \right\} 
\frac{\sin(\epsilon)}{\epsilon_0} \right] \ e^{-\frac{1}{2} E x^2} \nonumber \\
	    & \times & \sqrt{Q_3 - (Q_3 - Q_2) sn^{2}(\lambda \zeta,r)} \ 
	    e^{i\left(\theta(x)-Et\right)} 
\label{sol2}
\eea
\noindent The expression of $\theta(x)$ is obtained from Eq. (\ref{theta}). The qualitative behaviour
of the plots for identical $\mu_0(t)$ and condition on $\beta, \Gamma$ are the same
with that of Fig. 1. There is no power oscillation for vanishing $\beta, \Gamma$ as seen in Fig. 2(a).
The power-oscillation is seen in Fig. 2(b) for constant $\mu_0$ and ${\vert \beta \vert} > \Gamma$.
The solution grows without any upper bound for constant $\mu_0$ and ${\vert \beta \vert} = \Gamma$
as is evident from Fig. 2.(c). The management of instabilities for ${\vert \beta \vert} \leq \Gamma$
by choosing suitable $\mu_0(t)$ is shown in Figs. 2(d), 2(e) and 2(f).

\begin{figure}[H]
        \begin{subfigure}{0.32\textwidth}
                \centering
                \includegraphics[width=0.99\linewidth]{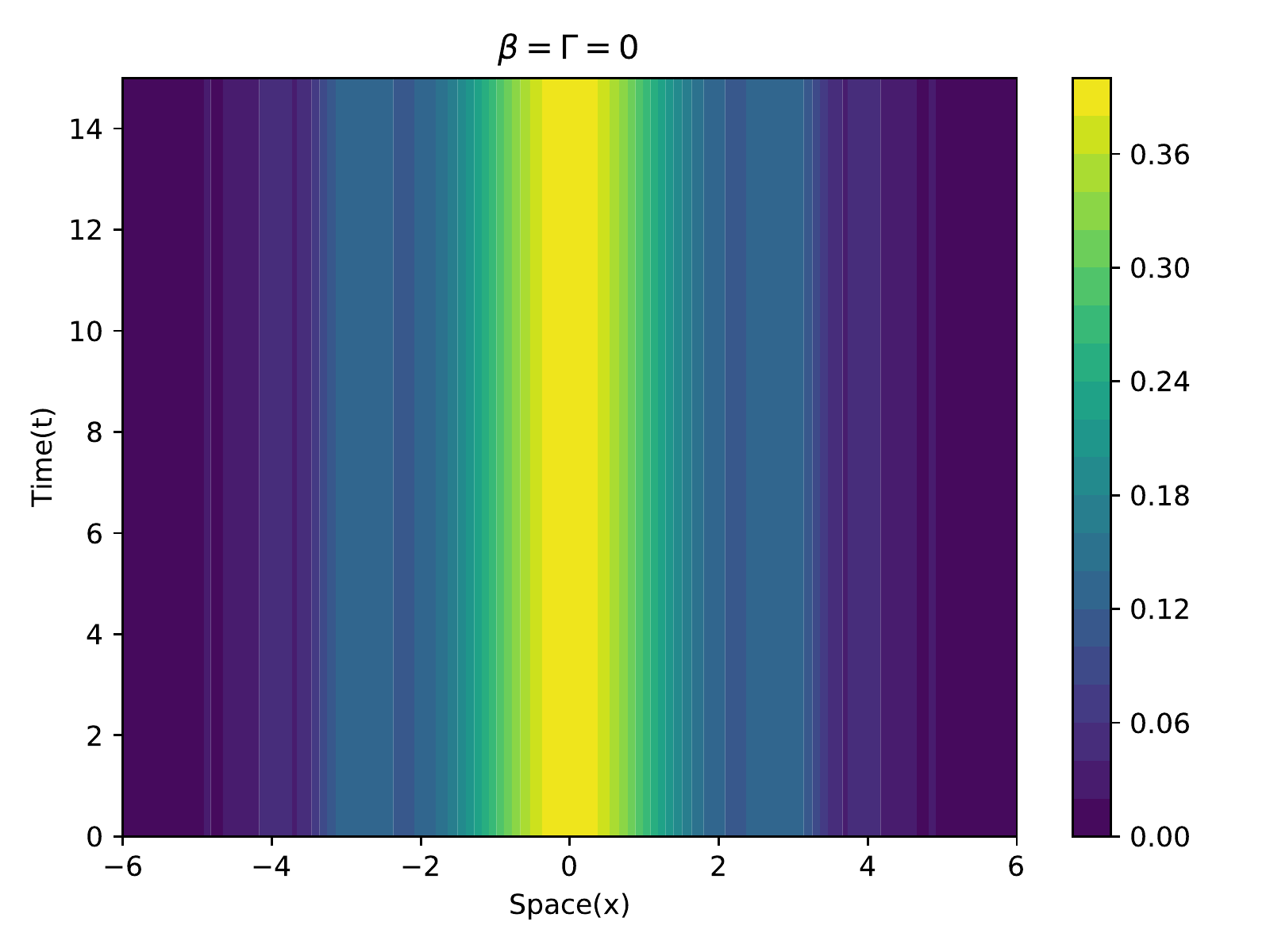}\quad
                \caption{}
                \label{fig:7}
        \end{subfigure}
        \begin{subfigure}{0.32\textwidth}
                \centering
                \includegraphics[width=0.99\linewidth]{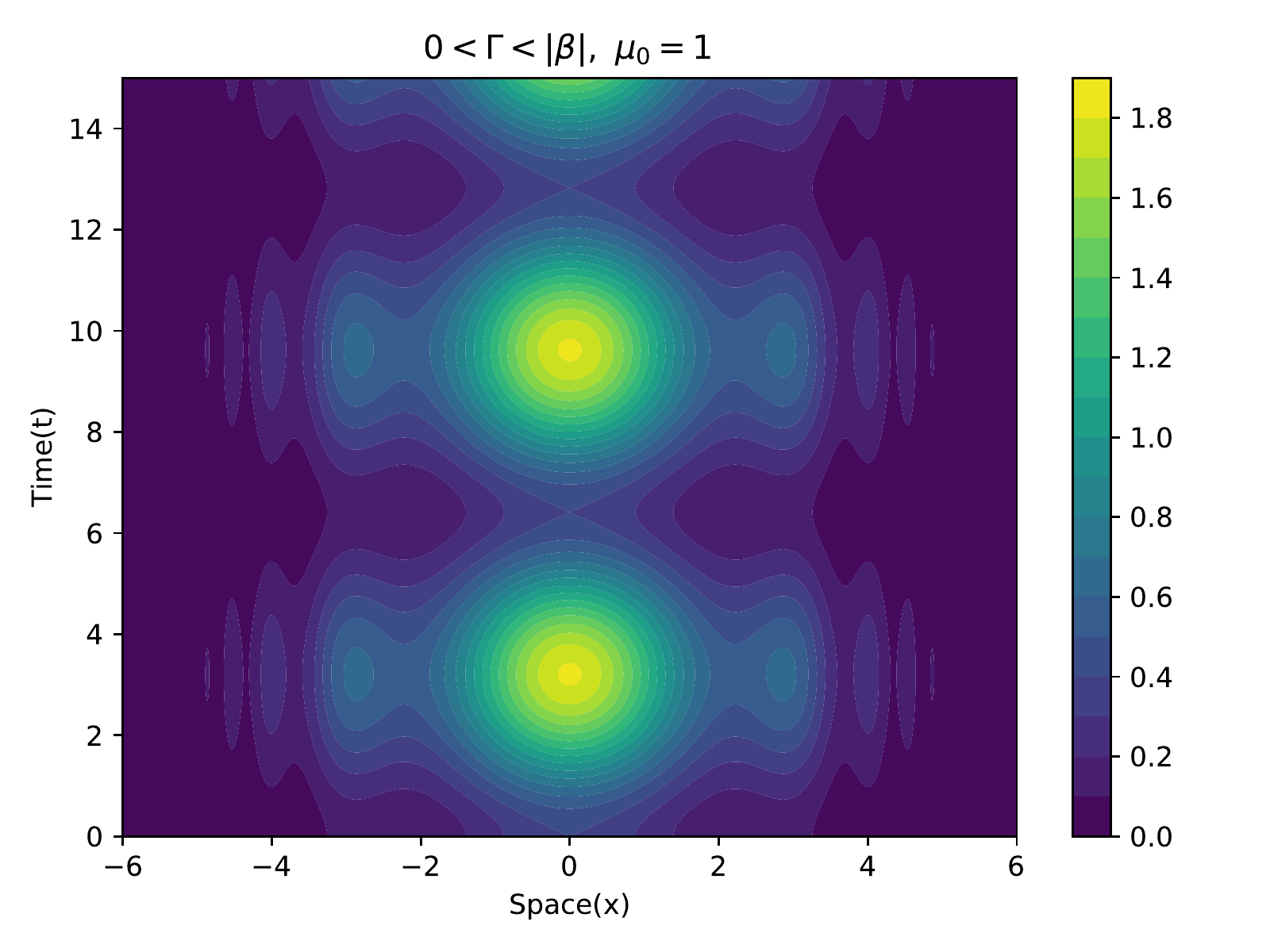}\quad
                \caption{}
                \label{fig:8}
        \end{subfigure}
        \begin{subfigure}{0.32\textwidth}
                \centering
                \includegraphics[width=0.99\linewidth]{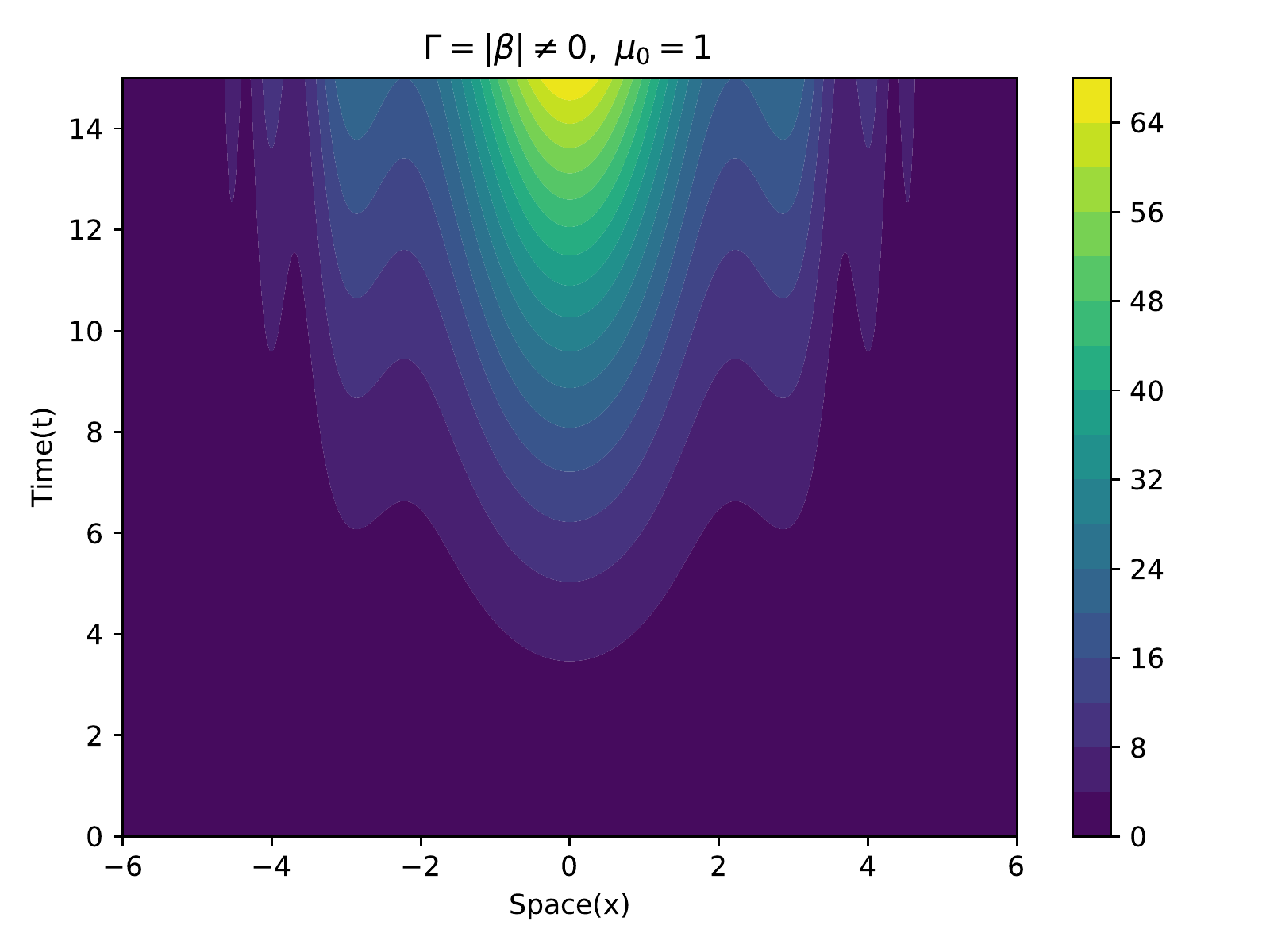}\quad
                \caption{}
                \label{fig:9}
        \end{subfigure}
        \medskip
        \begin{subfigure}{0.32\textwidth}
                \centering
                \includegraphics[width=0.99\linewidth]{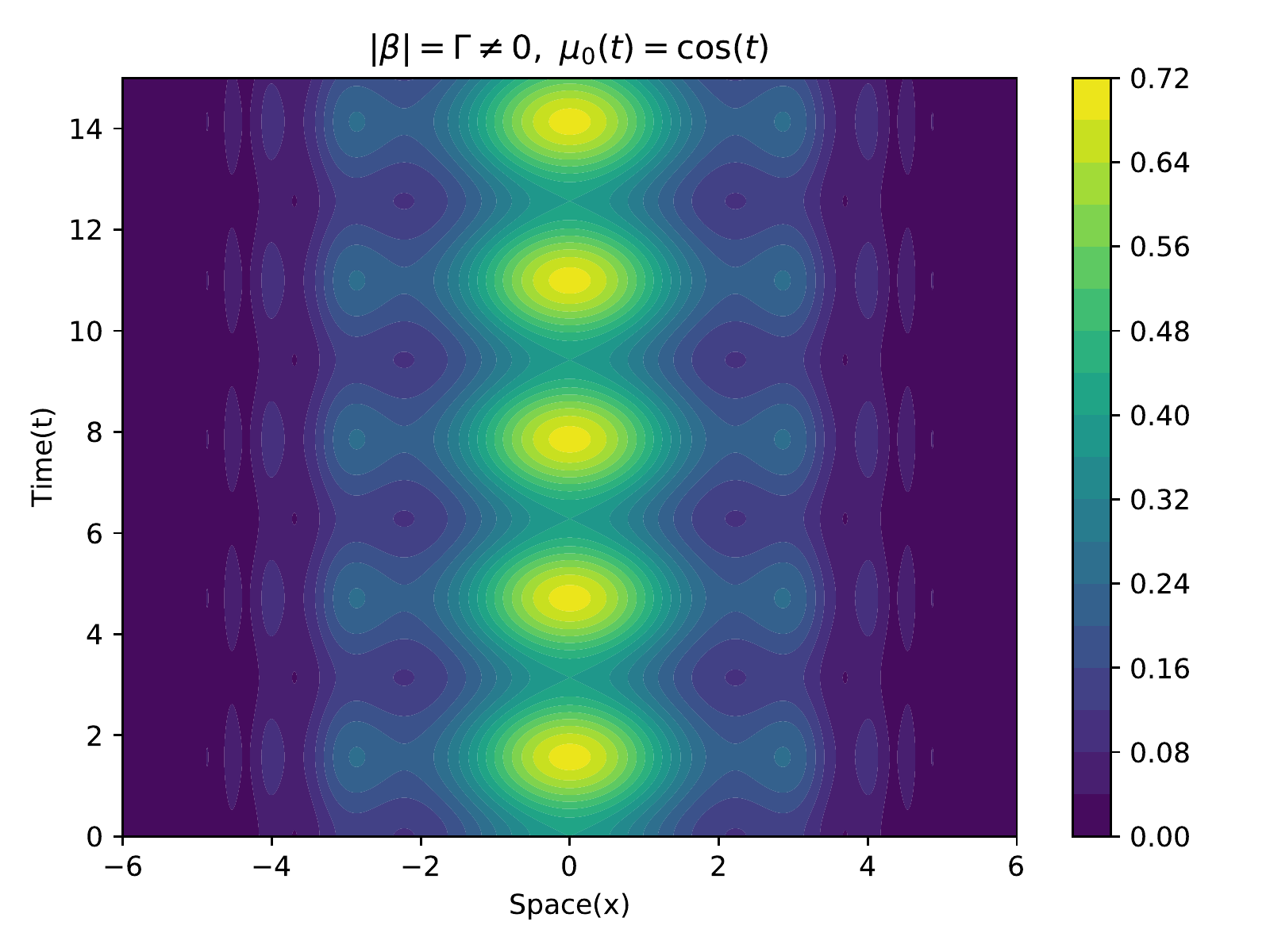}\quad
                \caption{}
                \label{fig:10}
        \end{subfigure}
        \begin{subfigure}{0.32\textwidth}
                \centering
                \includegraphics[width=0.99\linewidth]{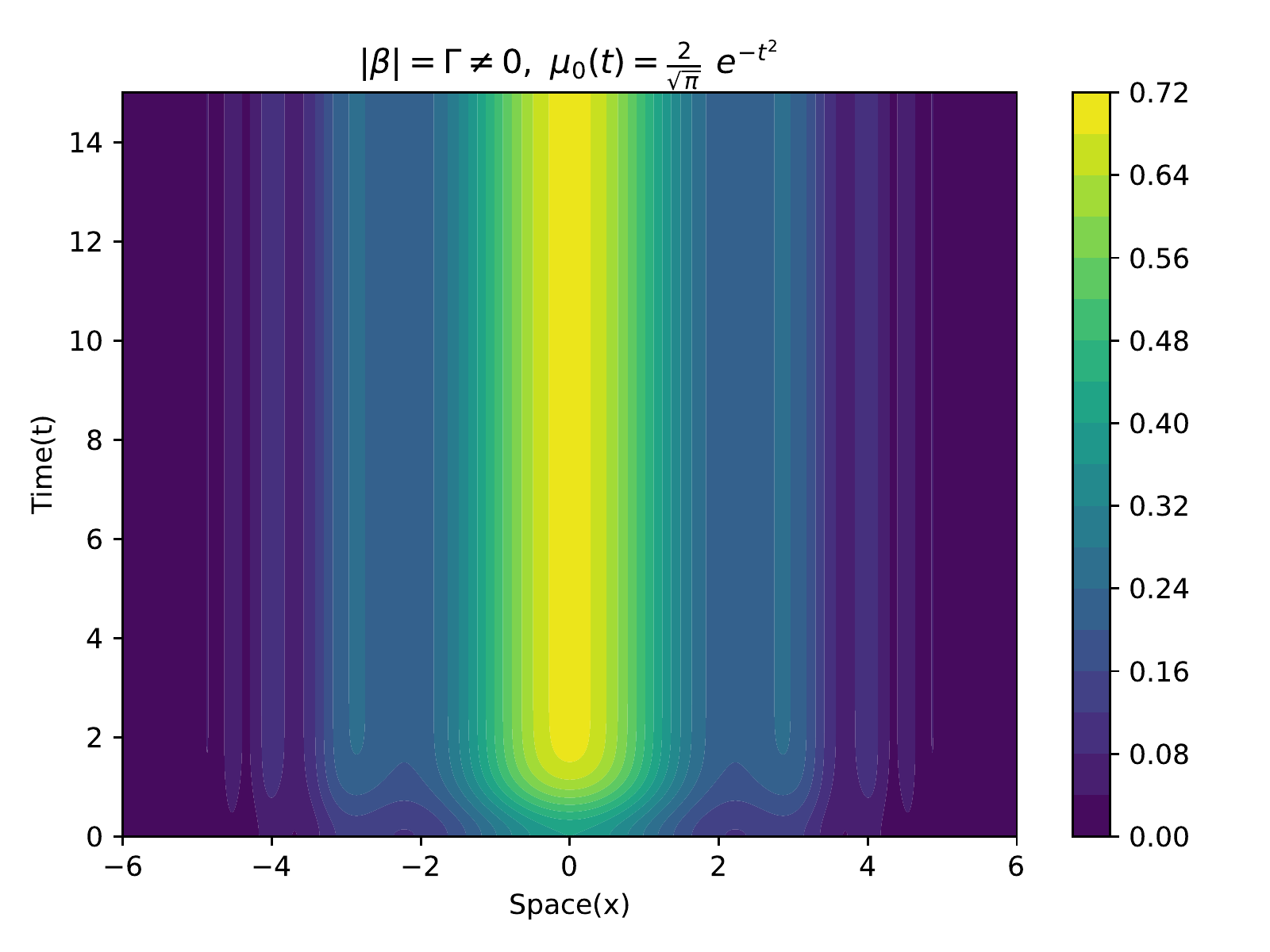}\quad
                \caption{}
                \label{fig:11}
        \end{subfigure}
        \begin{subfigure}{0.32\textwidth}
                \centering
                \includegraphics[width=0.99\linewidth]{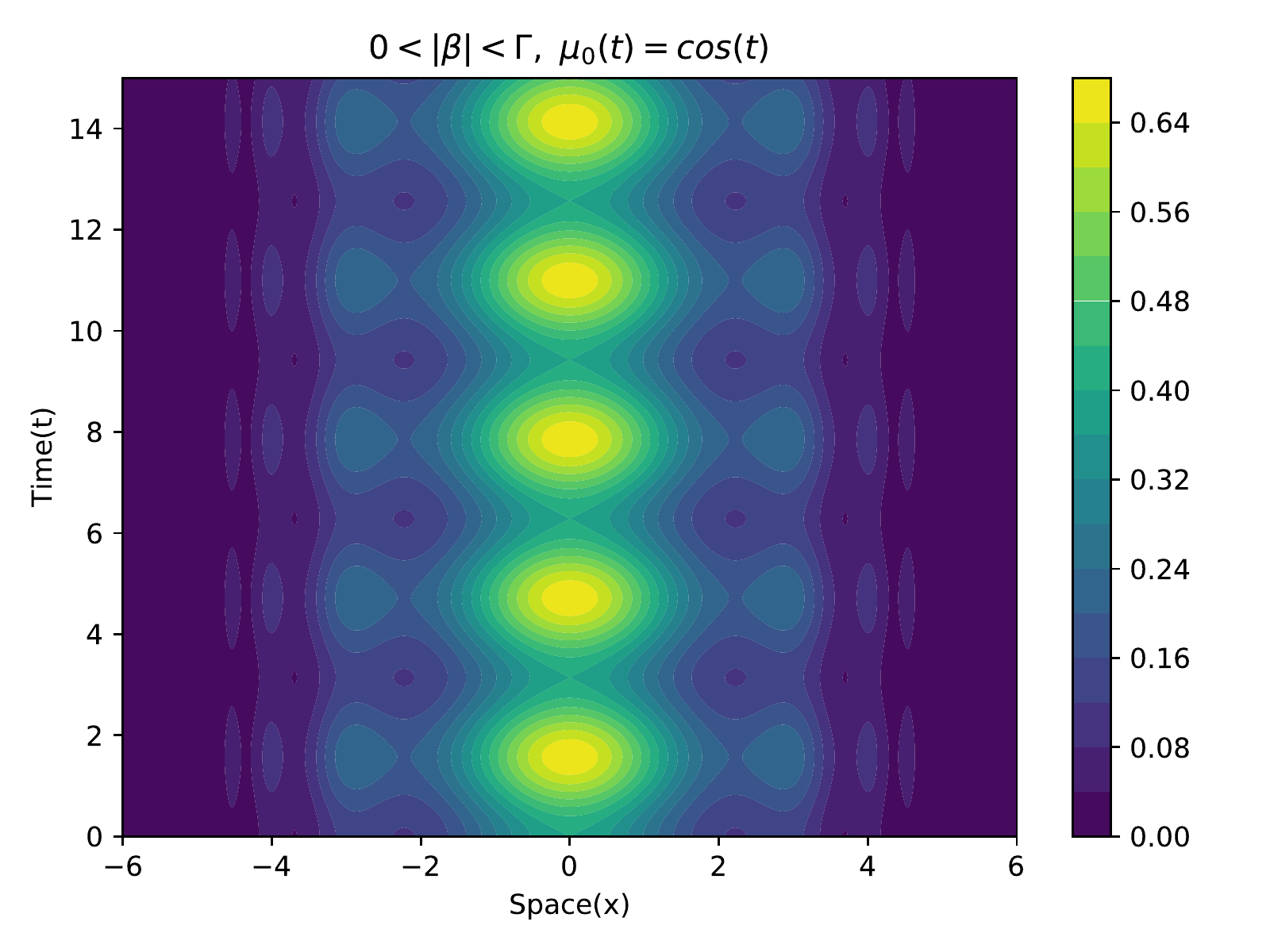}\quad
                \caption{}
                \label{fig:12}
        \end{subfigure}
	\captionsetup{justification=raggedright,singlelinecheck=false}
	\caption{ (Color online) Plot of Power i.e $\Psi^{\dagger}\Psi$ for $V(x) = E^2 x^2$ and the expression of 
	the components of $\Psi$ is given in Eq. (\ref{sol2}). Parameters: $W_1=W_2=1,\theta_1 = \frac{\pi}{6}, 
	\theta_2 = \frac{2\pi}{3}, \theta_3 = \frac{\pi}{3}, Q_1 = -0.3, Q_2 = 0.1, Q_3 = 0.2, \sigma = -1, 
	E = 0.12 $. In Fig(a): $\Gamma = |\beta| =0$, Fig(b): $\Gamma = 0.5, |\beta| =0.7, \mu_0 = 1$, 
	Fig(c): $\Gamma = |\beta|=0.5, \mu_0 =1$, Fig(d): $\Gamma=|\beta|=0.5,\mu_0(t) = cos(t)$,
	Fig(e): $\Gamma=|\beta|=0.5, \mu_0(t) = \frac{2}{\sqrt{\pi}} e^{-t^2}$, Fig(f): $\Gamma=0.5, 
	|\beta|=0.3,\mu_0(t) = cos(t)$ }
        \label{img2}
\end{figure}

\section{Systems with Complex $V(x)$}

The complex potentials appear in the context of optics through 
refractive index. The imaginary part of the refractive index
is introduced to describe the attenuation of propagating electromagnetic
waves in the given material. The ${\cal{PT}}$-symmetric complex potentials
have some interesting properties and have been studied extensively
in the context of NLSE\cite{Konotop1}. We present a few examples of complex
confining potential for NLSE with time-dependent LC and BLG, and space-time modulated
nonlinear strengths.

The nonlinear strengths $g_{ij}$ may be chosen to be constant,
purely time or space dependent or space-time modulated. One
notable aspect is that the constant and purely time-dependent
$g_{ij}$ may be considered for a complex confining potential, which
is not allowed for a real $V(x)$. The function $\theta(x)$ is determined
by Eq. (\ref{tt}) for a given $s(x)$ for constant $f(x)$ and $\rho$.
Such a freedom to consider a space-dependent $s(x)$ for constant $f$
is absent in Eq. (\ref{u1.5}) for the case of real potential. There is
a freedom in tailoring the nonlinear strengths $g_{ij}$ for a given $f(x)$
by choosing $f_1(x)$ and $f_2(x)$ appropriately, which may be exploited to
construct physically interesting systems.

{\bf Scarf II Potential:} The ${\cal{PT}}$-symmetric Scarf II complex potential has the form, 
\bea
V(x) = - \frac{V^2}{9} sech^2(x) - i V sech(x) \tanh(x)
\eea
\noindent The analytic and numerical solutions of one component NLSE with $V(x)$ has been discussed
in Ref. \cite{Shi,Musslimani}. We consider constant $f(x)= 2 \sigma$ for which $f_1$ and $f_2$ may
be chosen as constants or space-dependent subject to the condition $f(x)=\frac{1}{2} (f_1(x)+f_2(x))$.
We choose $ E = m = -1$ for simplicity and the phase $\theta(x)$ is determined from
Eq. (\ref{tt}) as $\theta(x) = \frac{V}{3} \arctan(\sinh(x))$.
We have to choose $\sigma < 0 $ so that Eqs. (\ref{theta0},\ref{R1}) and (\ref{u2.5}) are
consistent leading to the expression $R(x) = \frac{1}{\sqrt{|\sigma|}} \sech(x)$. 
The solution of Eq.(\ref{t-nlse1}) is obtained as,
\bea
\Phi = W {{|\sigma|}}^{-\frac{1}{2}} \sech(x) e^{i \big( {\frac{V}{3}} \arctan(\sinh(x)) + t \big)},
\eea
\noindent which describes a soliton. This completes the extension of the result to the
NLSE with time-modulated LC and BLG terms, and space-time modulated nonlinear strengths for the 
Scarf II potential. The expression of the components of $\Psi$ is given below,
\bea
\psi_1(x,t) &   =    & \left[ W_1 e^{i\theta_1} \cos(\epsilon) - \left\{ i |\beta| W_2
e^{i\left(\theta_2-\theta_3\right)} - \Gamma W_1 e^{i\theta_1} \right\}
\frac{\sin(\epsilon)}{\epsilon_0} \right] {{|\sigma|}}^{-\frac{1}{2}} \sech(x) \ 
e^{i \left( {\frac{V}{3}} \arctan(\sinh(x)) + t \right)} \nonumber \\
\psi_2(x,t) &  =     & \left[ W_2 e^{i\theta_2} \cos(\epsilon) - \left\{ i |\beta| W_1
e^{i\left(\theta_1+\theta_3\right)} + \Gamma W_2 e^{i\theta_2} \right\}
\frac{\sin(\epsilon)}{\epsilon_0} \right] {{|\sigma|}}^{-\frac{1}{2}} \sech(x) \ 
e^{i \left( {\frac{V}{3}} \arctan(\sinh(x)) + t \right)}
\label{sol3}
\eea
\noindent The qualitative behaviour of the plots for identical $\mu_0(t)$ and condition on $\beta, \Gamma$
are the same with that of Figs. 1 and 2. There is no power oscillation for vanishing $\beta, \Gamma$ as seen
in Fig. 3(a). The power-oscillation is seen in Fig. 3(b) for constant $\mu_0$ and ${\vert \beta \vert} > \Gamma$.
The solution grows without any upper bound for constant $\mu_0$ and ${\vert \beta \vert} = \Gamma$
as is evident from Fig. 3.(c). The management of instabilities for ${\vert \beta \vert} \leq \Gamma$
by choosing suitable $\mu_0(t)$ is shown in Figs. 3(d), 3(e) and 3(f).

\begin{figure}[!h]
        \begin{subfigure}{0.32\textwidth}
                \centering
                \includegraphics[width=0.99\linewidth]{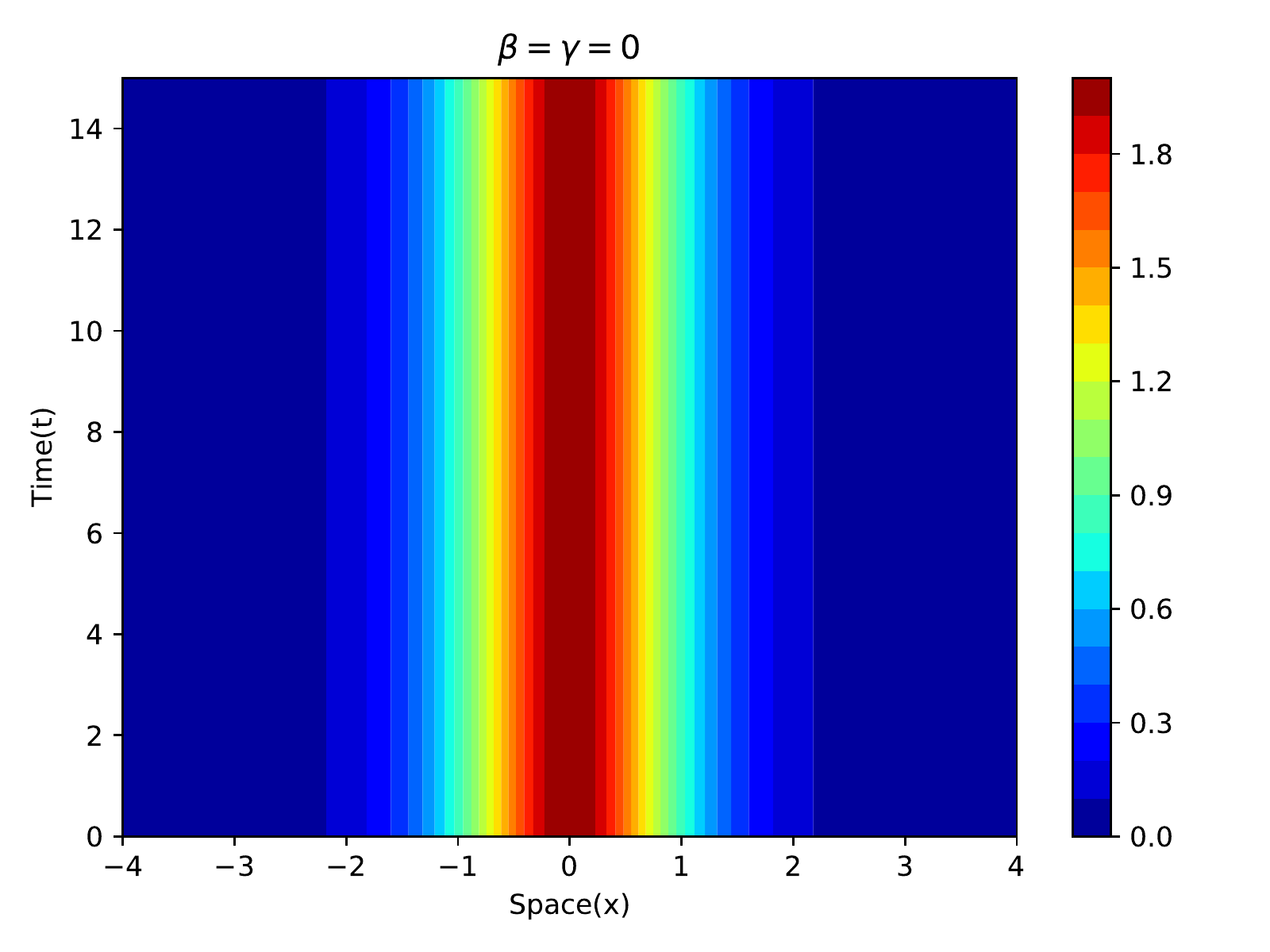}\quad
                \caption{}
                \label{fig:7}
        \end{subfigure}
        \begin{subfigure}{0.32\textwidth}
                \centering
                \includegraphics[width=0.99\linewidth]{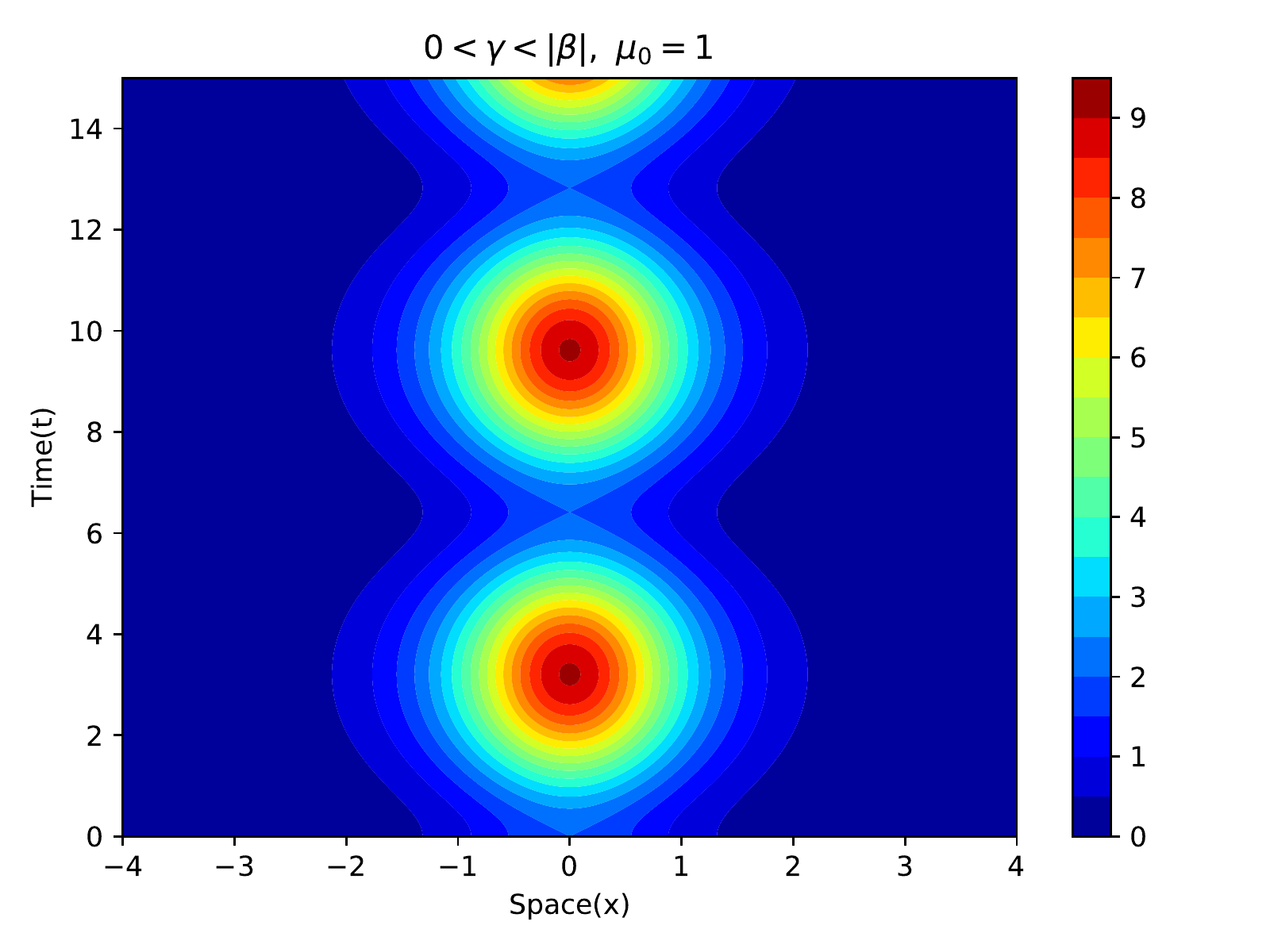}\quad
                \caption{}
                \label{fig:8}
        \end{subfigure}
        \begin{subfigure}{0.32\textwidth}
                \centering
                \includegraphics[width=0.99\linewidth]{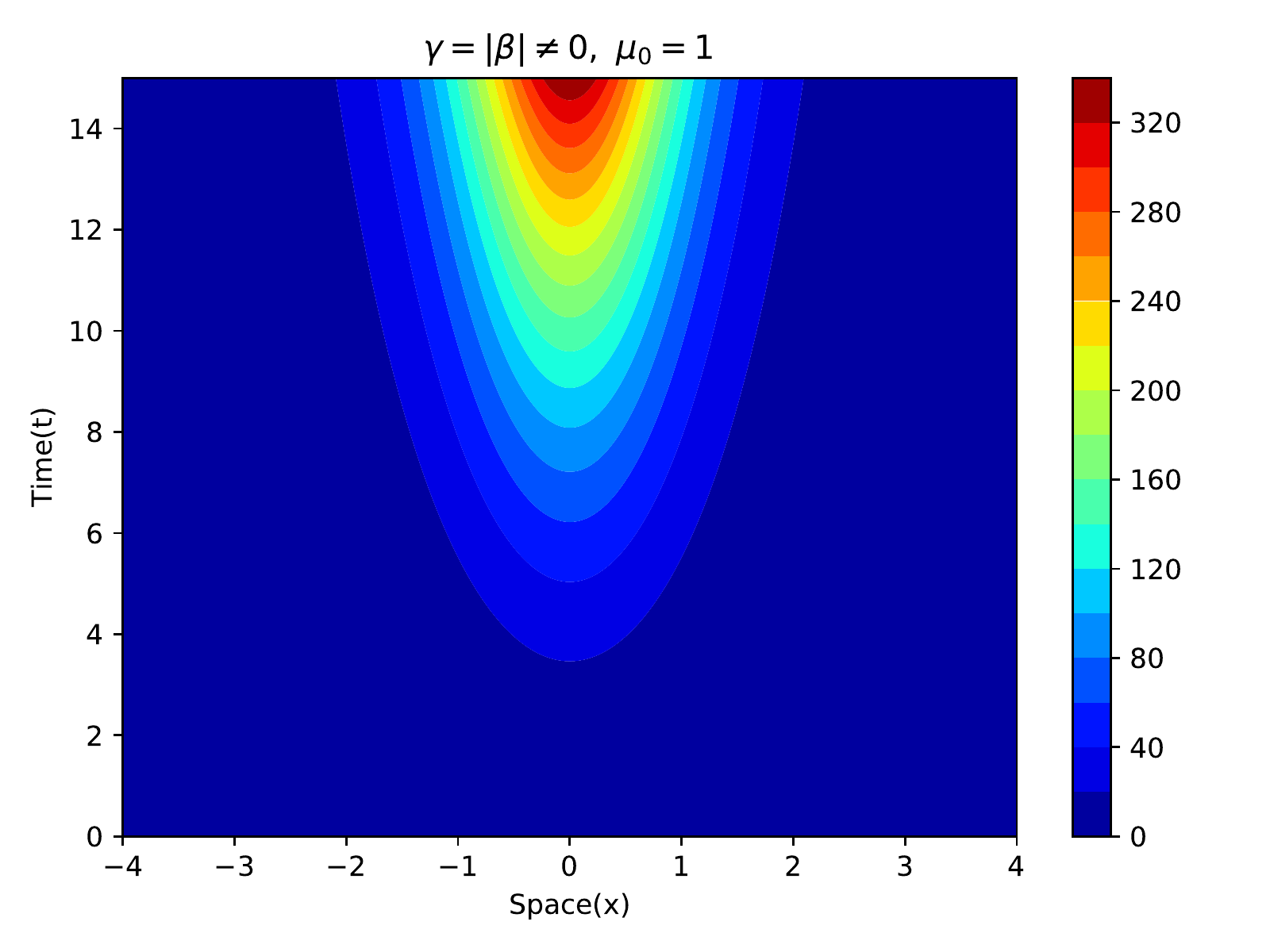}\quad
                \caption{}
                \label{fig:9}
        \end{subfigure}
        \medskip
        \begin{subfigure}{0.32\textwidth}
                \centering
                \includegraphics[width=0.99\linewidth]{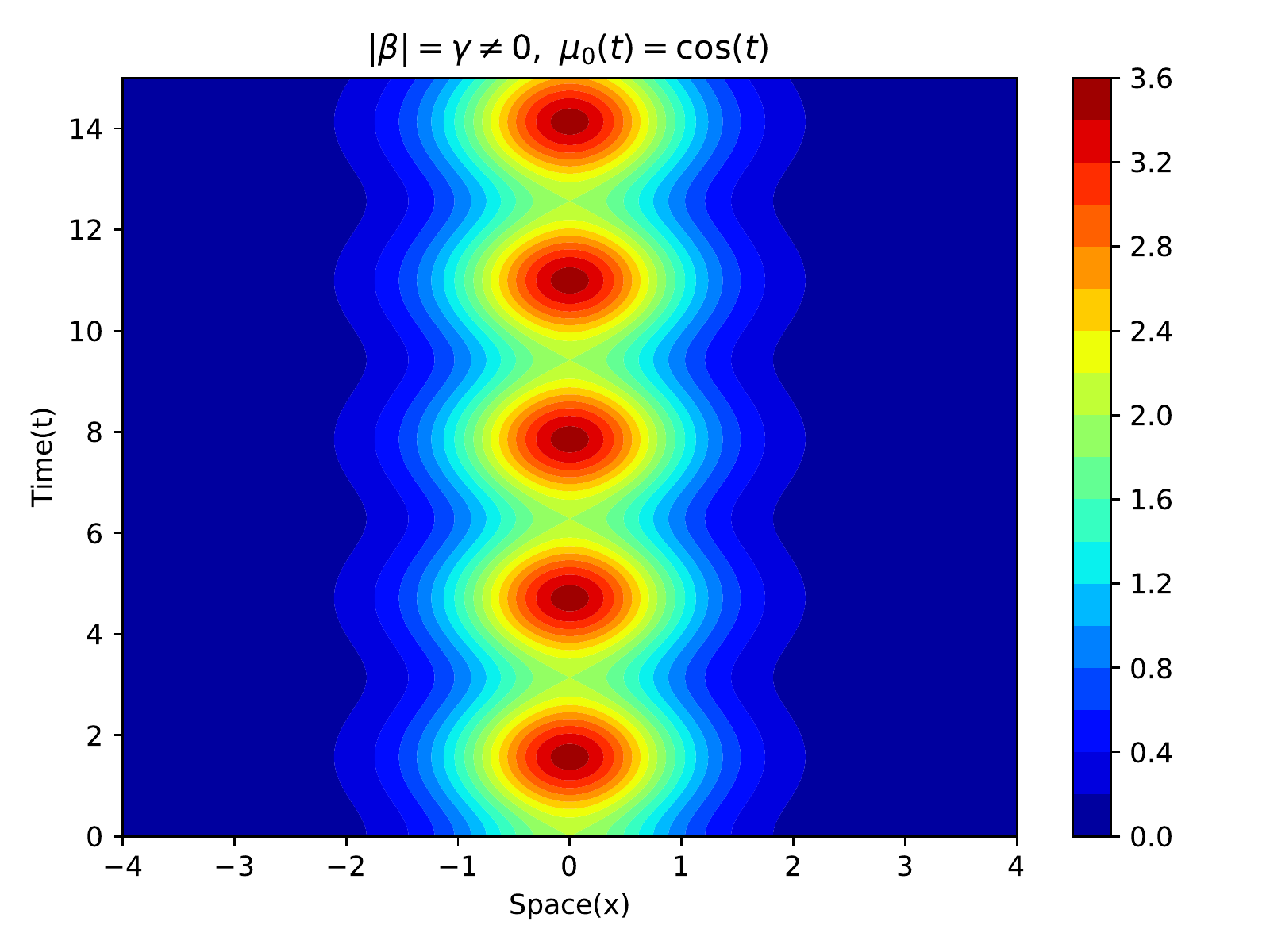}\quad
                \caption{}
                \label{fig:10}
        \end{subfigure}
        \begin{subfigure}{0.32\textwidth}
                \centering
                \includegraphics[width=0.99\linewidth]{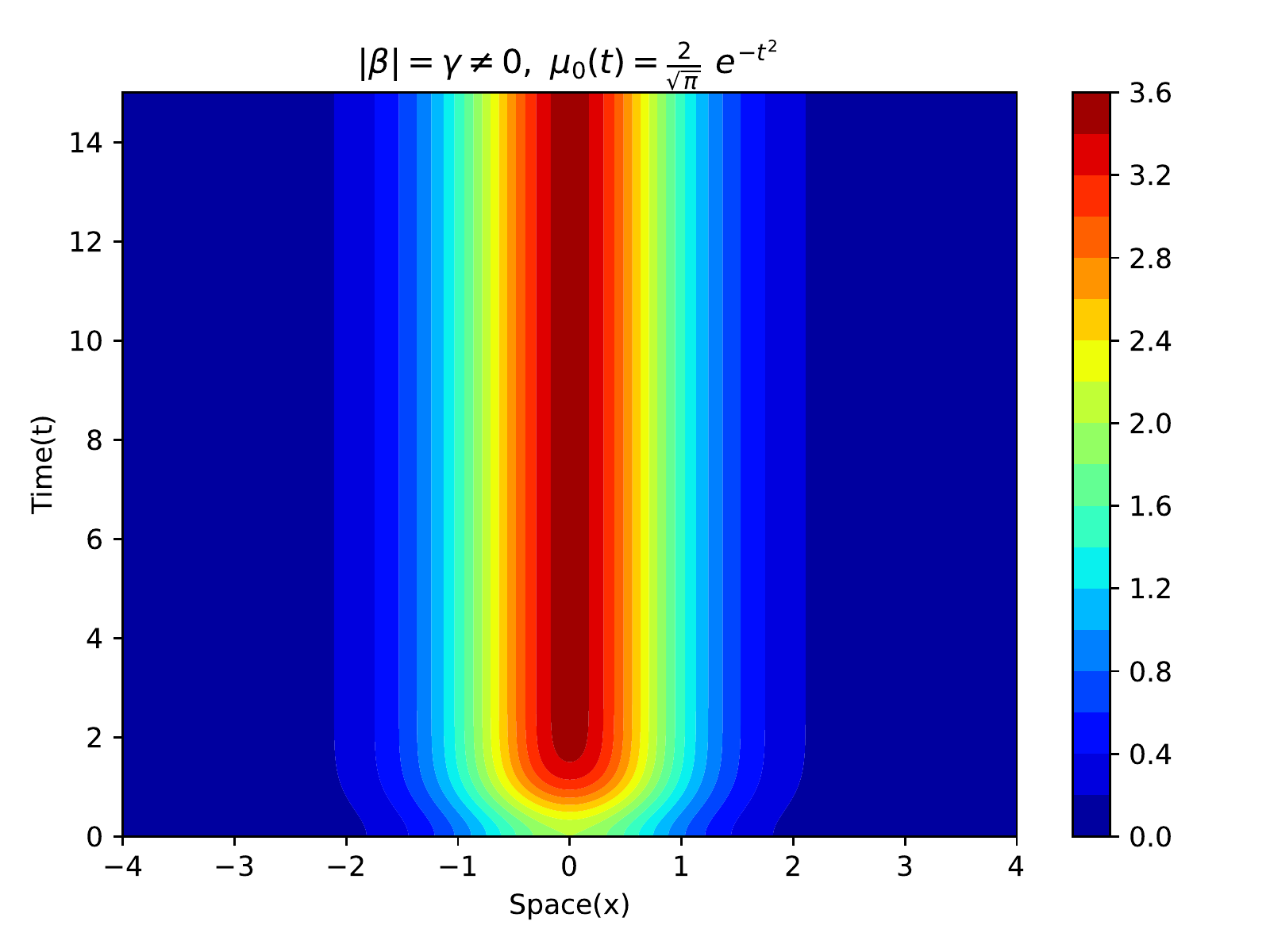}\quad
                \caption{}
                \label{fig:11}
        \end{subfigure}
        \begin{subfigure}{0.32\textwidth}
                \centering
                \includegraphics[width=0.99\linewidth]{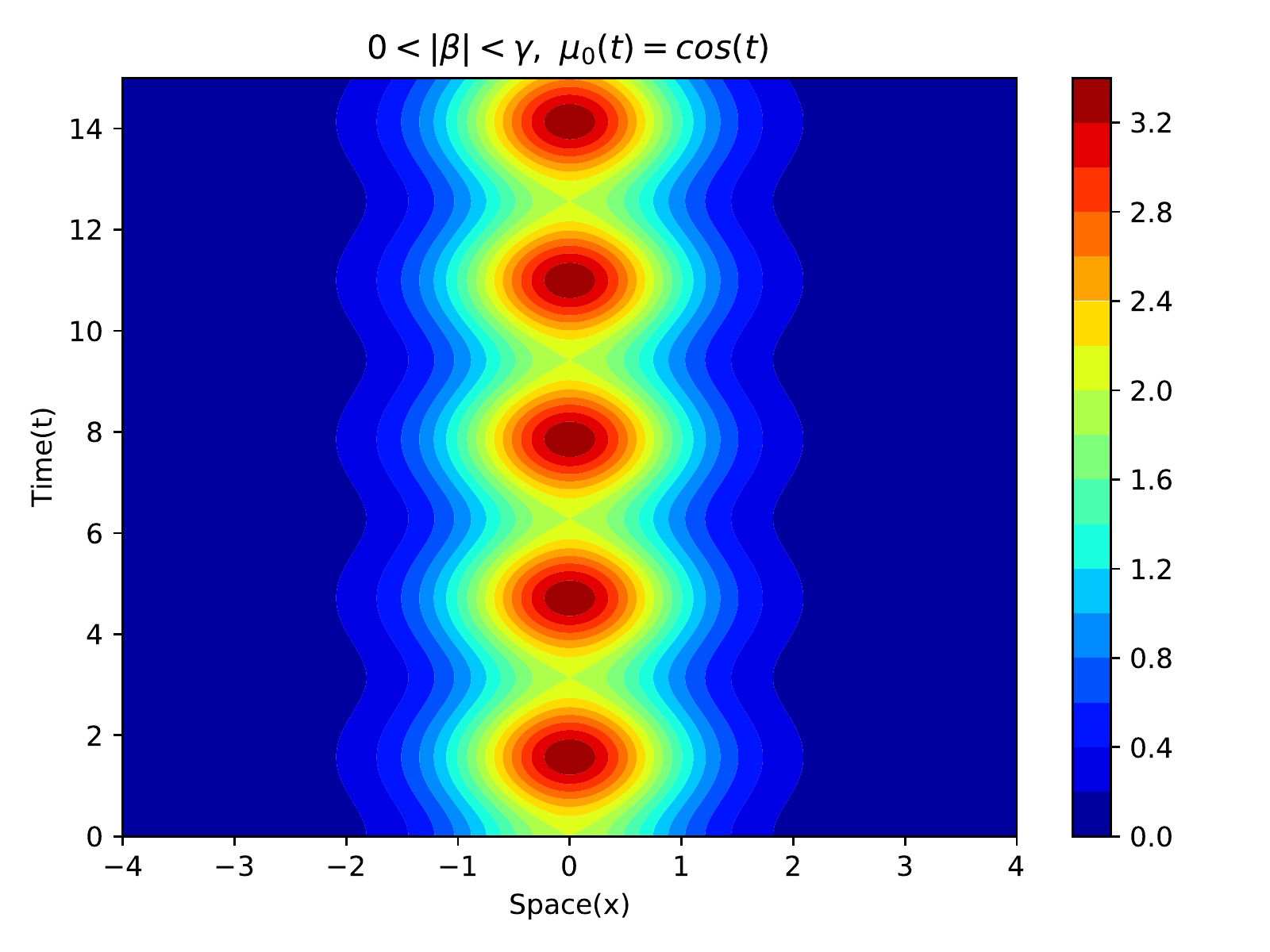}\quad
                \caption{}
                \label{fig:12}
        \end{subfigure}
	\captionsetup{justification=raggedright,singlelinecheck=false}
	\caption{ (Color online) Plot of Power i.e $\Psi^{\dagger}\Psi$ for Scarf-II potential and 
	the expression of the components of $\Psi$ in Eq. (\ref{sol3}). Parameters:
        $W_1=W_2=1,\theta_1 = \frac{\pi}{6}, \theta_2 = \frac{2\pi}{3}, \theta_3 = \frac{\pi}{3}, \sigma = -1$. 
	In Fig(a): $\Gamma = |\beta| =0$, Fig(b): $\Gamma = 0.5, |\beta| =0.7, \mu_0 = 1$, Fig(c): $\Gamma = |\beta|=0.5,
        \mu_0 =1$, Fig(d): $\Gamma=|\beta|=0.5,\mu_0(t) = cos(t)$, Fig(e): $\Gamma=|\beta|=0.5, \mu_0(t) 
        = \frac{2}{\sqrt{\pi}} e^{-t^2}$, Fig(f): $\Gamma=0.5,|\beta|=0.3,\mu_0(t) = cos(t)$}
        \label{img3}
\end{figure}

{\bf Periodic Potential} : The ${\cal{ PT}}$-symmetric complex potential,
\bea
 V(x) = - \cos^2(x) + i \sin(x)
\eea
\noindent has been considered in Ref. \cite{Musslimani} in the context of one component NLSE.
Following the procedure described above for $ E = m = 1$ and $ \sigma > 0$, the solution
is determined as,
\bea
\Phi = W \frac{1}{\sqrt{\sigma}} \sec(x) e^{i \big (\sin(x) - t \big)}.
\eea
\noindent We get exact solutions for NLSE with time-modulated LC and BLG terms,
and space-time modulated nonlinear strengths for the above periodic potential.

\subsection{Supersymmetry-inspired solution for $f(x) = 0$}

The techniques of supersymmetric quantum mechanics\cite{khare} may be used for $f(x)=0$
to construct a large number of complex potentials for which the NLSE with LC and BLG terms
are exactly solvable. The choice $f(x)$=0 can be implemented by fixing $f_2(x)=-f_1(x)$
and $f_1, f_2$ may be chosen to be either constants or space-dependent. There is a freedom
in choosing a large class of space-time modulated $g_{ij}$'s for a given complex $V(x)$.
The exact solutions of the NLSE are insensitive to the specific forms of $f_2=-f_1$ and $G(x,t)$.

The choice $f(x)=0$ reduces Eq.(\ref{R1}) to the linear Schr$\ddot{o}$dinger equation with an effective
potential $V_{eff}$:
\bea
R_{xx} + (E - V_{eff} ) R = 0, \ \ V_{eff} \equiv s(x)+\theta_x^2.
\label{R4.0}
\eea
\noindent We may fix $\theta(x)$ in terms of $\tilde{s}(x)$
such that Eqs. (\ref{theta0}) and (\ref{R1}) are consistent.
We define a function $h(x)$ which is related to $s(x)$ and $\theta(x)$ as follows,
\bea
s(x)  = - h_{x}, \ \ \theta(x) & = & \int^{x} h(x^{\prime}) dx^{\prime},
\label{s}
\eea 
\noindent for which $V_{eff}$ has the form of one of the supersymmetric partner potentials,
\bea
V_{eff} =   h^{2} - h_x .
\label{R4.1}
\eea
\noindent The function $h(x)$ is known as superpotential in
the context of supersymmetric quantum mechanics. The function
$R(x)$ plays the role of eigen-function and $E$ as the energy
eigen-value. The zero-energy eigen-function $R_0(x)$ has the
form,
\bea
R_0(x) & = & N_0 \ e^{-\theta(x)},
\label{R4.2}
\eea
\noindent where $N_0$ is a normalization constant. The expression of $R(x)$
for $E \neq 0$ can also be obtained for shape-invariant potentials\cite{khare}.
We restrict our discussions for $E=0$ in this article and $R_0(x)$ can be
determined for any $h(x)$ independent of whether it corresponds to shape-invariant
potential or not. The factorization of the effective potential $V_{eff}$ as in Eq. (\ref{R4.1})
ensures exact solution for $R_0(x)$. The consistency of Eqs. (\ref{theta0}) and (\ref{R1})
fixes the imaginary part and the complex potential has the expression,
\bea
V(x) = -h_x(x) + i \left [ h_x(x) - 2 h^2(x) \right ].
\eea
\noindent The superpotential corresponding to all the known examples of solvable
supersymmetric quantum system may be used to construct complex potential $V(x)$
and the corresponding solution for $R(x)$. The solution $R(x)$ corresponding to
these potentials describe localized nonlinear modes.
The expression for $\Phi(x,t)$ is determined as,
\bea
\Phi(x,t)= N_0 W e^{-\theta(x) + i \theta(x) }.
\eea
\noindent The important point to note is that
corresponding to each exactly solved quantum mechanical problem by using supersymmetry,
the corresponding superpotential may be used to find a complex potential for which
exact localized nonlinear modes are obtained.

We present a few examples to complement the general discussions by including expressions for
$h(x), V(x)$ and $\Phi(x)$. Note that $\Phi(x,t)$ is time-independent, since we have
chosen $E=0$ for the presentation of results:\\
{\bf {Polynomial potential}} :
\bea
&& h(x)= \omega_0 + \omega_1 x + \omega_2 x^3, \ (\omega_0, \omega_1, \omega_2) \in \ \mathbb{R}^+,\nonumber \\
&& V(x)= -(1-i) \left ( \omega_1 + 3 \omega_2 x^2 \right ) - 2 i  \left (\omega_0 +
\omega_1 x +\omega_2 x^3 \right )^2,\nonumber \\
&& \Phi(x)= N_0 W e^{-(1-i) \left ( \omega_0 x + \frac{\omega_1}{2} x^2 + \frac{\omega_2}{4} x^4 \right ) }\nonumber .
\eea
\noindent The real part of the potential describes a harmonic oscillator, while the imaginary  part is a
sextic potential. The function $\Phi(x)$ is localized in space even for $\omega_2=0$ for which the real part of the potential
is constant and the imaginary part is given by an quadratic potential.

{\bf{Exponential Potential}} :
\bea
&& h(x) =  A - B e^{-ax} , \ (a, A, B) \in \mathbb{R}^+, \nonumber \\
&& V(x)= - B a e^{-a x} + i [ 2 A^2 - B (a + 4 A)  e^{-a x} + 2 B^2 e^{-2 a x} ], \nonumber \\
&& \Phi(x) = N_0 W \ exp[ - \frac{B}{a} e^{-ax} - A x] \ e^{i\big( \frac{B}{a} e^{-a x} + A x \big)}.\nonumber 
\eea
\noindent The real part of the $V(x)$ describes a potential well of depth $Ba$ and $\Phi(x)$ is localized in space.

{\bf Systems with singular phase}: The phase singularity in two and higher dimensional optical systems has
interpretation in terms of vortices, wavefront dislocation, etc.\cite{berry1}. We present two examples
exhibiting phase singularity in one dimension.  The first example corresponds to the superpotential for a system
describing quantum particle in a box of length $L$:
\bea
&& h(x) = -\frac{\pi}{L} cot(\frac{\pi x}{L}),\nonumber \\
&& V(x) =  - \frac{\pi^2}{L^2} cosec^2(\frac{\pi x}{L}) + i [ \frac{\pi^2}{L^2} \big( 1 - cot^2(\frac{\pi x}{L})  \big)], \nonumber \\
&& \Phi(x) = W \ \sqrt{\frac{2}{L}} \ \sin(\frac{\pi x}{L}) \  e^{-i \  ln(\sin(\frac{\pi x}{L}))}.
\eea
\noindent The nonlinear mode is localized within the box and vanishes at the boundary. However, the phase is singular
at the boundaries of the box. The second example corresponds to the superpotential of Rosen-Morse potential:
\bea
&& h(x) =  N \tanh(x), \ N \in \mathbb{R}^+, \nonumber \\
&& V(x)  =  - N sech^{2}(x) - i [ 2 N^2 - N (2N + 1) sech^2(x)], \nonumber \\
&&\Phi(x) = W sech^{N}(x) exp[-i \big( N ln(sech(x)) \big) ].\nonumber
\eea
\noindent Both the the real and imaginary parts of $V(x)$ define potential-wells of finite depth. The potential is
well-defined on the whole line. The nonlinear mode `$\Phi(x)$ is localized in space.
The phase of $\Phi(x)$ is singular for ${\vert x \vert} \rightarrow \infty$.

\section{Discussion and Conclusion}

We have investigated exact solvability of a class of vector NLSE with time-modulated LC and BLG terms,
and space-time modulated cubic nonlinear terms in presence of an external complex potential. We have
taken a two-step approach to find exact solutions. The BLG and LC terms are removed completely through
a non-unitary transformation at the cost of modifying the time-modulation of the nonlinear strength.
In general, the real-valued nonlinear interaction becomes complex after the non-unitary transformation.
Further, the method of non-unitary transformation is applicable only if LC and BLG terms have identical
time-modulation.  The separation of time and space co-ordinates as well as real and
imaginary parts of the equation requires the nonlinear strengths to be time-independent and real-valued.
This is achieved by fixing the nonlinear strengths appropriately such that the method of separation
works. The time-dependence of the nonlinear strengths is determined in terms of the time-modulation
of LC and BLG terms along with an arbitrary function $G(x,t)$, while the space-dependence may be chosen
in terms of two arbitrary functions $f_1(x), f_2(x)$ and $G(x,t)$. 

In the second step, the resulting equations are analyzed by using the method
of similarity transformation which involves writing the differential
equation in a new co-ordinate system and multiplying the amplitude by a space-dependent
scale-factor. The treatment for analyzing solvability for real and complex potentials
are different. The space-time dependence of the power $P$ is factorised in terms of
the product of space-dependent and time-dependent functions. The time-dependence
of $P$ solely depends on the form LC and BLG couplings and independent
of the specific form of external potential or the strengths of the space-time modulated
cubic nonlinearity. However, the space-dependence of $P$ is determined in terms of
the choice of the external potential as well as space-modulation of the nonlinear strengths.

We have constructed several examples of exactly solvable models for
constant, purely time dependent,  purely space dependent and space-time
dependent nonlinear-strengths for vanishing external potential. This has been
done for constant as well as time-modulated LC and BLG terms. On the other hand,
the method employed in this article allows to construct 
exact solutions for the non-vanishing real potential for purely space dependent
or space-time modulated nonlinear strengths. Exact solutions for constant
or purely time dependent nonlinear strengths can not be constructed
by using the method. The complex external potential allows more flexibility
in constructing exactly solvable models. In particular, exactly solvable models for
constant, purely time dependent,  purely space dependent and space-time
dependent nonlinear-strengths, and time-modulated LC and BLG terms are constructed.
One interesting result is that exact localized nonlinear modes with spatially constant
phase may be obtained for any real $V(x)$ for which the corresponding
linear Schr$\ddot{o}$dinger equation is solvable.
Further, for the case of complex potential, we have developed a method based on
supersymmetric quantum mechanics to construct several exactly solvable models. In fact,
corresponding to each exactly solved quantum mechanical problem by using supersymmetry,
the corresponding superpotential may be used to find a complex potential for which
exact localized nonlinear modes are obtained. We find a few complex potentials for
which exact nonlinear modes exhibit singular phases.

A few notable features independent of specific models are the following:
\begin{itemize}
\item The exact solutions do not depend at all on the choice of the
function $G(x,t)$. The reasons may be attributed to the particular ansatz
chosen for the separation of variables, and real and imaginary parts of
the differential equation. It is possible that new solutions depending on
$G(x,t)$ may be found for the system. Such solution, if exists,
have to be found numerically or by different analytical methods.

\item The exact solutions do not depend on the specific choices
of $f_1(x)$ and $f_2(x)$, but, on their average $f(x)$.
This may again be attributed to the specific ansatz chosen
for finding exact solutions. However, analytical
and/or numerical methods may be employed to verify whether
or not the solutions are insensitive to the choice of specific
$f_1(x)$ and $f_2(x)$.

\item The power-oscillation in time is absent for time-independent strength of nonlinear
terms. This holds for constant as well as time-modulated LC and BLG terms. The result
is to be contrasted with the system considered in Ref. \cite{pkg}, where
power-oscillation is seen for constant LC, BLG and nonlinear strength. There is
no contradiction, since the nonlinear term in Eq. (\ref{nlse1}) for $g_{ij}$ given
by Eq. (\ref{G1}) can not be cast in the form $(\psi^{\dagger} M \psi)^{2}$ such that
$A$ is pseudo hermitian with respect to $M$. The power oscillation is observed
in ref. \cite{pkg} for constant LC, BLG and nonlinear term whenever $A$ is $M$ pseudo hermitian.
Thus, the power-oscillation for constant LC and BLG terms in \cite{pkg} may be attributed to
the specific form of the nonlinear interaction.

\item The case $f(x)=0$ for which exact localized nonlinear modes are obtained for real as well as
complex potentials is one of the salient features of the class of NLSE considered in this article.
The method based on supersymmetric quantum mechanics to construct localized nonlinear
modes for complex potential needs to be explored further for its applicability
to other types of NLSE with external potential. 

\end{itemize}

The system defined by Eq. (\ref{nlse1}) is directly relevant in the context of optics. The NLSE has
been studied through approximate and/or numerical methods previously for constant
nonlinear strengths and constant or specific time-modulated LC and BLG terms. We have
given a generic framework in which a class of exact solutions may be found for various
combinations of space-time modulated nonlinear strengths, time-modulated LC and BLG terms and external
potential. Specific realizations of some of these models in realistic physical scenario may enrich the
current understanding on the subject.

\begin{acknowledgments}

The work of PKG is supported by a grant (SERB Ref. No. MTR/2018/001036) from the Science \& Engineering 
Research Board(SERB), Department of Science and Technology, Govt.of India, under the MATRICS scheme. 
SG acknowledges the support of DST INSPIRE fellowship of Govt.of India(Inspire Code No. IF190276).

\end{acknowledgments}

\section{Appendix-I: Solutions of a nonlinear equation}

We present the solutions of the following equation:
\bea
u_{\zeta \zeta} + \omega^2(\zeta) u - \frac{C^2}{u^3} - 2 \sigma u^3 = 0, 
\label{u}
\eea
\noindent where $\sigma$ and $C$ are real constants. We discuss two special cases before embarking
on the general solutions:\\
(I) ${\bf \sigma = 0}$: Eq. (\ref{u}) describes Ermakov-Pinney equation. The general solution is given by,
\bea
u(\zeta) = \sqrt{v_1^2 + \frac{C^2}{W^2} v_2^2}, \ W(\zeta)= v_1 v_{2,\zeta} - v_2 v_{1,\zeta},
\eea
\noindent where $v_1$, $v_2$ are two independent solutions of the equation $v_{\zeta \zeta}+ \omega^2(\zeta) v=0$ satisfying
the initial conditions at $\zeta=\zeta_0$: $v_1(\zeta_0)=u(\zeta_0), v_{1,\zeta}(\zeta_0) = u_{\zeta}(\zeta_0),
v_2(\zeta_0) \neq 0, v_{2,\zeta}(\zeta_0)=0$. We have used the convention $v_{i,\zeta}=\frac{dv_i}{d\zeta}, i=1,2$.
For constant $\omega$, the choice $v_1=\sin(\omega \zeta), v_2=\cos(\omega \zeta)$ leads to the periodic solution,
\begin{eqnarray}
u(\zeta) & = & \sqrt{\sin^2(\omega \zeta)+\frac{C^2}{w^2}\cos^2(\omega \zeta)}.
\end{eqnarray}
\noindent For a periodic $\omega(\zeta)$ with period $T$, we obtain Hill's equation which has been studied extensively
in the literature\cite{hill}. The solutions for a generic periodic $\omega(\zeta)$ with the condition
$v_1(0) =0, v_{1,\zeta}=1, v_2(0)=1$ are stable for ${\vert v_1(T) + v_{2,\zeta}(T) \vert} < 2$\cite{hill}.\\
(II) \ ${\bf C = 0}$: In Eq.(\ref{u}) we replace $\omega^{2}(\zeta)$ with constant $m$ which can take both 
positive and negative values. This describes a NLSE and it's solution is well known,
\begin{eqnarray}
u(\zeta) & = & \sqrt{\frac{m}{\sigma}} \ \sec \big(\sqrt{m} \zeta \big), \ (m, \sigma) > 0 \nonumber \\
  & = & \sqrt{\frac{|m|}{|\sigma|}} \ sech \big(\sqrt{|m|}\zeta \big), \ (m, \sigma ) < 0.
\end{eqnarray}
III. \ ${\bf C \neq 0, \sigma \neq 0, \omega^2 \equiv m=}\textrm{\bf Constant}$:
The transformation $u(\zeta) = \sqrt{Q(\zeta)}$ along with a change in the variable from $\zeta$ to $y = \sqrt{|\sigma|} \zeta$
followed by integration transforms Eq. (\ref{u}) in the following form:
\bea
Q_y^{2} = \sgn(\sigma) \ 4 \ (Q-Q_1) (Q-Q_2) (Q-Q_3)
\label{Q2}
\eea
\noindent where $Q_1, Q_2, Q_3$ are the roots of the cubic equation $Q^3 - \frac{m}{\sigma} Q^2 + \frac{C _1}{\sigma} Q
- \frac{C^2}{\sigma}=0$, $C_1$ is an integration constant and $\sgn(\sigma)$ is the signum function. The expressions for
the roots in terms of $\omega, \sigma, C, C_1$ are too lengthy and will not be presented here. However, for given values
of $Q_i$'s in a particular solution, corresponding values of  $\omega, \sigma, C, C_1$ may be
obtained by using the properties of the roots:
$Q_1 Q_2 Q_3 = \frac{C^2}{\sigma}, \ \ Q_1Q_2+Q_2Q_3+Q_1Q_3 = \frac{C_1}{\sigma},
Q_1+Q_2+Q_3 =  \frac{m}{\sigma}$.
The real solutions of Eq. (\ref{Q2}) satisfy $(Q-Q_1) (Q-Q_2) (Q-Q_3) \geq 0$ for $\sigma >0$, while
$(Q-Q_1) (Q-Q_2) (Q-Q_3) \geq 0$ for $\sigma <0$. The finite and stable solutions of Eq.(\ref{Q2})
are presented below based on boundedness of the solution $Q$ in terms $Q_i$. A particular ordering among thee
roots are considered for presentation of the results following the discussions in Ref. \cite{Beitia}:\\
(a) $ {\bf \sigma > 0, 0 < Q_1 \leq Q \leq Q_2 < Q_3} $ : The solution of the Eq.(\ref{Q2}) reads,
\bea
Q = Q_1 + (Q_2 - Q_1) sn^2[\lambda y, r],
\label{+Q1}
\eea
\noindent where $\lambda = \sqrt{Q_3 - Q_1}$ and $ r^2 = {\frac{Q_2 - Q_1}{Q_3 - Q_1}}$. The solution can be expressed
in terms of hyperbolic function in the limit of $Q_3 \rightarrow Q_2$:
\bea
Q = Q_1 + (Q_2 - Q_1) \tanh^2[\lambda y]
\label{+Q2}
\eea

(b) ${\bf  \sigma > 0,  0 < Q_1 < Q_2 < Q_3 \leq Q} $: The solution of Eq.(\ref{Q2}) reads,
\bea
Q = Q_1 + \frac{Q_3 - Q_1}{sn^2(\lambda y, r)},
\eea
\noindent where $\lambda^2 = Q_3 - Q_1$ and $r^2 = \frac{Q_2 - Q_1}{Q_3 - Q_1}$.
The solution reduces to an elementary singular periodic function for $Q_1=Q_2$, while it gives a singular soliton
if $Q_3 = Q_2$.

(c) ${\bf \sigma <0, Q_1 < 0 < Q_2 \leq Q \leq Q_3}$: In this case, we get real finite stable solution.
The solution for $\sigma < 0$ has the form:
\bea
Q = Q_3 - (Q_3 - Q_2) sn^2(\lambda y, r)
\label{-Q1}
\eea
\noindent where $ \lambda ^{2} = Q_3 - Q_1$ and $ r^2 = \frac{Q_3 - Q_2}{Q_3 - Q_1}$.

\end{document}